\documentclass[12pt]{article}
\usepackage{amsfonts}
\usepackage{amsmath, amssymb, amsthm, latexsym, eepic }
\usepackage[dvips]{color}
\usepackage{graphicx, subfigure}
\usepackage{psfrag}
\usepackage{mathrsfs}
\usepackage{wrapfig}
\usepackage{color}

\newcommand{\beq}{\begin{equation}}
\newcommand{\eeq}{\end{equation}}
\newcommand{\beqa}{\begin{eqnarray}}
\newcommand{\eeqa}{\end{eqnarray}}
\newcommand{\CR}{\nonumber \\}

\renewcommand{\k}{\kappa}

\newcommand{\trace}{{\rm Tr}}

\newcommand{\beeq}{\begin{eqnarray}}
\newcommand{\eeeq}{\end{eqnarray}}

\def\C{{\mathbb{C}}} 
\def\R{{\mathbb{R}}} 
\def\Z{{\mathbb{Z}}} 

\def\rem#1{}

\newcommand{\bel}{\begin{eqnarray}}
\newcommand{\ee}{\end{eqnarray}}

\renewcommand{\title}[1]{\vbox{\center\LARGE{#1}}\vspace{5mm}}
\renewcommand{\author}[1]{\vbox{\center\large#1}\vspace{5mm}}

\numberwithin{equation}{section}

\begin{document}

\begin{titlepage}

\bigskip
\hfill\vbox{\baselineskip12pt
\hbox{IPMU18-0106}
}

\begin{center}
\Large{ \bf
Equivariant $U(N)$ Verlinde algebra \\
from Bethe/Gauge correspondence
}
\end{center}
\bigskip
\bigskip

\begin{center}
\large 
Hiroaki Kanno$^{a,b,}$\footnote{kanno@math.nagoya-u.ac.jp}, 
Katsuyuki Sugiyama$^{c,}$\footnote{sugiyama@scphys.kyoto-u.ac.jp}, and
Yutaka Yoshida$^{d,}$\footnote{yutaka.yoshida@ipmu.jp} \\
\bigskip
\bigskip
$^a${\small {\it Graduate School of Mathematics, Nagoya University,
Nagoya, 464-8602, Japan}}\\
$^b${\small {\it KMI, Nagoya University,
Nagoya, 464-8602, Japan}} \\
$^c${\small {\it Department of Physics, Kyoto University, Kyoto, 606-8502, Japan}} \\
$^d${\small {\it Kavli IPMU (WPI), UTIAS, The University of Tokyo, Kashiwa, Chiba 277-8583, Japan.}}
\end{center}

\bigskip
\bigskip

\begin{abstract}
We compute the topological partition function (twisted index) of $\mathcal{N}=2$ $U(N)$ Chern-Simons theory
with an adjoint chiral multiplet on $\Sigma_g \times S^1$. The localization technique shows 
that the underlying Frobenius algebra is the equivariant Verlinde algebra which is obtained from 
the canonical quantization of the complex Chern-Simons theory regularized by $U(1)$ equivariant parameter $t$. 
Our computation relies on a Bethe/Gauge correspondence which allows us to represent 
the equivariant Verlinde algebra in terms of the Hall-Littlewood polynomials $P_\lambda(x_B, t)$ with a specialization 
by Bethe roots $x_B$ of the $q$-boson model. We confirm a proposed duality to 
the Coulomb branch limit of the lens space superconformal index of four dimensional $\mathcal{N}=2$ theories
for $SU(2)$ and $SU(3)$ with lower levels. In $SU(2)$ case we also present more direct computation 
based on Jeffrey-Kirwan residue operation.
\end{abstract}
\end{titlepage}

\setcounter{footnote}{0}
\newpage
\baselineskip=18pt

\tableofcontents

\section{Introduction}

Localization method in the computation of supersymmetric (SUSY)  gauge theories 
(\cite{Pestun:2016zxk} and references therein) 
gives exact results which are useful for confirming non-perturbative 
dualities and correspondence among the partition functions and correlation
functions. It also reveals a relation to the integrable system such as Bethe/Gauge
correspondence, which was first observed in \cite{Moore:1997dj} 
and later developed in \cite{Nekrasov:2009uh, Nekrasov:2009ui, 
Nekrasov:2009rc, Nekrasov:2014xaa}. A kind of \lq\lq mother\rq\rq\ theory of 
non-trivial dualities involving low dimensional SUSY theories is 
the six dimensional $\mathcal{N}=(2,0)$ superconformal field theory
that arises as a low energy effective world-volume theory of $M5$ branes.
Namely we can make use of the fact that twisted\footnote{To keep supersymmetry in lower dimensions, 
it is often necessary to make a twist along the curved compactification manifold.
This twist should be distinguished from another twist which is used to obtain topological theory 
from the resulting supersymmetric theory.}  compactification of the six dimensional theory 
on a manifold $M_n$ of dimension $n$ gives a SUSY gauge theory
in $(6-n)$ dimensions, which is a source of otherwise unexpected correspondence 
between the field theory on $M_n$ and the SUSY gauge theory.
One of the most intriguing examples is  AGT(W) correspondence \cite{Alday:2009aq, Wyllard:2009hg},
where twisted compactification on a punctured Riemann surface $\Sigma_{g,n}$
gives four dimensional $\mathcal{N}=2$ superconformal theories of class $\mathcal S$
and the instanton partition function of the class $\mathcal S$ theory computes conformal blocks on $\Sigma_{g,n}$.
The theory of our interest in this paper is the compactification on a three manifold $M_3$
denoted as $T[M_3]$. With an appropriate twisting we can keep $\mathcal{N}=2$ supersymmetry 
and the theory $T[M_3]$ gives supersymmetric theory
on complementary 3-manifold $\widetilde M_3$. In spite of interesting proposal of 
so-called 3d-3d correspondence \cite{Terashima:2011qi, Terashima:2011xe, Dimofte:2011ju, Dimofte:2011py}, 
the theory $T[M_3]$ for general three manifold is only partly explored.

However, if $M_3$ is a Seifert manifold which is an $S^1$ bundle over a Riemann surface $\Sigma$, 
we have good chances for getting detailed information about $T[M_3]$, since $S^1$-compactification of 
$\mathcal{N}=(2,0)$ theory gives 5 dimensional super Yang-Mills theory which is relatively tractable. 
For example, it is known that complex Chern-Simons theory
 (Chern-Simons theory with a complexified gauge group $G_{\C}$) 
is obtained by a compactification of 6d theory on the (squashed) lens space $L(\kappa,1)_b$
 \cite{Cordova:2013cea, Lee:2013ida, Dimofte:2014zga, Dimofte:2016pua}. 
 The Chern-Simons theory with a compact gauge group $G$
 is a renowned example of topological quantum field theory (TQFT) \cite{Witten:1988hf}. 
 After the canonical quantization on $\widetilde M_3 = \Sigma \times S^1$
 with the periodic time along $S^1$ which gives the trace, 
 the partition function gives a two dimensional TQFT on $\Sigma$ that
 counts the dimensions of physical Hilbert space due to the vanishing TQFT Hamiltonian. 
 If one introduces the Wilson loop operators along the time direction, they create
 punctures on $\Sigma$ and the Hilbert space is mathematically 
 identified with the space of conformal blocks of WZNW model.  
 Hence,  the Verlinde algebra or the fusion ring of the current algebra
 \cite{Verlinde:1988sn, Gepner:1990gr, Intriligator:1991an,  Witten:1993xi}
underlies the two dimensional TQFT from the Chern-Simons theory. 
 Quite similarly complex Chern-Simons theory is also 3 dimensional TQFT and 
 the quantization on $\Sigma \times S^1$ gives a 2d TQFT \cite{Gukov:2015sna, Andersen:2016hoj}.
Furthermore, in \cite{Gukov:2016lki} by considering six dimensional $\mathcal{N}=(2,0)$ superconformal field theory 
on $L(\kappa,1) \times \Sigma \times S^1$, it is proposed that 2d TQFT obtained 
from complex Chern-Simons theory or $\mathcal{N}=2$ Chern-Simons theory with an adjoint chiral multiplet 
on $\Sigma \times S^1$ has a dual description in terms of the Coulomb branch limit of the superconformal 
index on the lens space $L(\kappa,1)$. When $G_{\mathbb{C}}=SL(N, \mathbb{C})$, the superconformal index of our concern is associated with the
class $\mathcal{S}$ theory of $A_{N-1}$ type obtained by the compactification on $\Sigma$. 
The family of these indices defines 2d TQFT \cite{Gadde:2009kb, Benini:2011nc}
and according to the general principles of 2d TQFT, 
the Coulomb branch limit of the lens space index can be evaluated by gluing those associated 
with three punctured sphere\footnote{The $A_{N-1}$ type class $\mathcal{S}$ theory associated 
with $\Sigma_{0,3}$ is called $T_N$ theory. Except for $N=2$ it does not allow Lagrangian description 
in general.} $\Sigma_{0,3}$.
In this paper we  check this proposal by explicitly computing the genus $g$ partition functions
on the Chern-Simons theory side. 

The powerful localization technique for $\mathcal{N}=2$ SUSY theory on Seifert manifold was first
worked out in \cite{Ohta:2012ev} and further elaborated by \cite{Benini:2015noa, Benini:2016hjo, Closset:2016arn}. 
When the manifold is $\Sigma \times S^1$, we can make the partition function localized on discrete
SUSY vacua (critical points of superpotential) which coincide with the solutions (Bethe roots)
to the Bethe ansatz equation of integrable lattice models called 
phase model and $q$-boson model \cite{Okuda:2012nx, Okuda:2013fea, Okuda:2015yea}. 
By supersymmetric localization
the path integral of topologically twisted Chern-Simons-matter theory reduces to infinite magnetic sum 
and multi-contour residue integrals called Jeffrey-Kirwan (JK) residues \cite{Benini:2013xpa}. 
One might hope that one can check the duality to the superconformal  index of the class $\mathcal{S}$ theory
by summing up infinitely many JK residues in the localization formula. 
Unfortunately, there is a crucial subtlety in evaluating the partition function (topologically twisted index)\footnote{
In this paper we call topologically twisted index of Chern-Simons-matter theory simply (topological) partition function.}
on higher genus Riemann surfaces. When the genus is larger than one and the gauge group is non-abelian,  
the projective condition of the JK residues is violated due to the one-loop determinant of vector multiplet. 
Hence, JK residues are ill-defined and naive residue operation does not reproduce lens space index. 
Therefore we need an alternative method to evaluate the topological partition function on higher genus surfaces.
In this paper, we employ 2d TQFT viewpoint and quantum integrable structure 
behind Chern-Simons-matter theory a.k.a Bethe/Gauge correspondence. 
Since the Chern-Simons-matter theories are topologically twisted along Riemann surface, 
we expect that the Chern-Simons-matter theories possess the structure of 2d TQFT.
As we summarized in Appendix A, in 2d TQFT the partition function on higher genus Riemann surface is reconstructed 
by genus zero correlation functions.  Although JK residues for genus zero case are well-defined, 
it is technically difficult to evaluate infinitely many JK residues in practice, for example $SU(3)$ theories.

 If the magnetic sum is performed before the JK residue operation and integration contours are deformed 
 to enclose saddle points of effective twisted superpotential, the topological partition function and correlation functions are 
 given by finite summations over solutions of the saddle point equations.  
 But it is still hard to evaluate them, because it is usually impossible to solve the saddle point equations explicitly.  
To overcome such a difficulty, in this paper we use a combinatorial algorithm to evaluate correlation functions
without knowing explicit form of solutions. This algorithm features the Hall-Littlewood polynomials $P_\lambda(x,t)$
that arise naturally from the algebraic Bethe ansatz of the $q$-boson model \cite{Korff:2013rsa}. 
In $U(N)$ theories, the saddle point equations agree with Bethe ansatz for $N$ particle sector of $q$-boson model,
while the number of sites corresponds to the level $\k$ of the Chern-Simons theory. 
Thus, the Bethe/Gauge correspondence helps us to compute the partition function of 2d TQFT 
on Chern-Simons theory side, whose algebraic structure (the deformed Verlinde algebra) 
is related to the algebra of Hall-Littlewood polynomials on the space of Bethe roots.
We emphasize that it is $U(N)$ Chern-Simons theories that are related to the $q$-boson model, 
or the Hall-Littlewood polynomials.
However, once the topological partition function and correlation functions of 
$U(N)= (U(1) \times SU(N))/\Z_N$ theories are given, 
those of $SU(N)$ theories are obtained by decomposing $U(N)$ theory to $U(1)$ part and $SU(N)$ part.   
We show that the twisted indices of $SU(N)$ theories reproduce Coulomb branch limit of lens space index for $SU(2)$ and 
$SU(3)$ with lower levels, which confirms the proposal in \cite{Gukov:2016lki}.
We also provide a result for level 2 $U(4)$ theory, but there is no corresponding computation 
on the superconformal index side at the moment.

\subsection{Complex Chern-Simons theory and $\mathcal{N} =2$ Chern-Simons theory with adjoint matter}

Compactification of 6d theory on the (squashed) lens space
\beq
L(\kappa,1)_b := \{ (z,w) \in \C^2 ; b^2 |z|^2 + b^{-2} |w|^2 =1 \} / \Z_{\kappa},
\label{lens}
\eeq
has been shown to give a complex Chern-Simons theory\footnote{For $G=U(N), SU(N)$ the complexified gauge group
is $G_{\C} = GL(N,\C), SL(N, \C)$. In this paper we only consider these cases.}
\cite{Cordova:2013cea, Lee:2013ida, Dimofte:2014zga, Dimofte:2016pua}.
The orbifold action in \eqref{lens} is defined by $(z,w) \mapsto (e^{2\pi i/\kappa} z, e^{-2\pi i/\kappa}w)$.
The action of complex Chern-Simons theory is
\beq
\mathcal{S} = \frac{q}{8\pi} \int_{M_3} \trace \left( \mathcal{A} \wedge d \mathcal{A}
+ \frac{2}{3} \mathcal{A} \wedge \mathcal{A}\wedge \mathcal{A} \right)
+ \frac{\bar q}{8\pi} \int_{M_3} \trace \left( \bar{\mathcal{A}} \wedge d \bar{\mathcal{A}}
+ \frac{2}{3} \bar{\mathcal{A}} \wedge \bar{\mathcal{A}} \wedge \bar{\mathcal{A}}\right),
\eeq
where $\mathcal{A} =  \mathbf{A} + i \Phi$ is a complex gauge field and $q = \k + i \sigma$ is 
the complex coupling constant.  
For the invariance under the large gauge transformations the real part of the coupling $\kappa$ has to be integer. 
Under the complex gauge transformation both $ \mathbf{A}$ and 
$\Phi$ transform as a gauge field. But if the gauge transformation is restricted to be
real, $\Phi$ transforms as a matter in the adjoint representation. Or the pair
$( \mathbf{A}, \Phi)$ is regarded as coordinates on the cotangent bundle of the space of
connections for the compact gauge group $U(N)$ or $SU(N)$. 
When we consider the compactification on $L(\kappa,1)_b$, the imaginary part of the Chern-Simons coupling 
is related to the squashing parameter by $\sigma = \k \frac{1- b^2}{1+ b^2}$ \cite{Cordova:2013cea, Lee:2013ida}.

In this paper we only consider the case $b=1$\footnote{When $b=1$ we denote the lens space simply by  $L(\kappa,1)$.}, 
namely $\sigma =0$. The action on $\Sigma \times S^1$ becomes
\beq
S^{\sigma=0}  = \frac{\k}{4\pi} \int_{\Sigma \times S^1} \trace \left( A \wedge D_0 A  + 2 A_0 \wedge (dA + A \wedge A )
-2 \phi_0 \wedge d_A \phi - \phi \wedge D_0 \phi \right),
\eeq
where we have made a decomposition $\mathbf{A} = A + A_0 dx_0 , \Phi = \phi + \phi_0 dx_0$ 
and $x_0$ is a coordinate along $S^1$.
Since there are no time derivatives of $A_0$ and $\phi_0$ we obtain
\beq
F_A - \phi \wedge \phi = 0, \qquad   d_A \phi = 0 \label{flat}
\eeq
as constraints for the Hilbert space of the canonical quantization of complex Chern-Simons theory.
In fact \eqref{flat} is the flatness condition of the total curvature $\mathcal{F} := d \mathcal{A} + \mathcal{A} \wedge \mathcal{A}$
restricted on $\Sigma$. A crucial fact which connects the complex Chern-Simons theory and 
the twist of $\mathcal{N}=2$ super Chern-Simons theory with an adjoint matter multiplet is the fact
that
\beq
\mathcal{M}_H := \{ F_A - \phi \wedge \phi = 0,~~d_A \phi= 0 \} / \mathcal{G}_{\C}
=  \{ F_A - \phi \wedge \phi = 0,~~d_A \phi= d_A^\dagger \phi = 0 \} / \mathcal{G},
\label{Hitchin}
\eeq
which gives an equivalence of a holomorphic and a Hermitian description of the Hitchin moduli space\footnote{
Precisely speaking we have to impose some stability condition in the holomorphic description.}. 
In \eqref{Hitchin} $\mathcal{G}_{\C}$ is the group of complex gauge transformations,  
while $\mathcal{G}$ is that of real gauge transformations. 
It is the relations in the second description of the Hitchin moduli space that arise naturally 
as the equations of motion (topological gauge fixing conditions) 
in the topological twist of $\mathcal{N}=2$ Chern-Simons theory with an adjoint matter multiplet \cite{Gukov:2015sna, Okuda:2015yea}.
In the second description we impose the additional condition $d_A^\dagger \phi = 0$ 
in compensation for the reduced gauge symmetry $\mathcal{G}$.
An important role of this additional condition is
that this allows us to introduce $SO(2)$ rotation acting on the space components $(\phi_1, \phi_2)$ of the one form $\phi$  \cite{Gukov:2015sna}. 
In fact, the equation $d_A \phi = 0$ is invariant under the $SO(2)$ rotation only when it is combined with 
the condition $d_A^\dagger \phi = 0$.

On the other hand the adjoint chiral multiplet $\phi = \phi_1 + i \phi_2$ in the $\mathcal{N}=2$ Chern-Simons theory 
is originally a complex scalar field with $U(1)$ flavor symmetry. But the $R$-symmetry of
the 3 dimensional $\mathcal{N}=2$ SUSY algebra is $U(1)_R$ 
and there is a freedom of $U(1)_R$ charge assignment $r$ for $\phi$.
Since the topological twist of 3d theory on $\Sigma \times S^1$  is a redefinition of 
2d local Lorentz symmetry $SO(2)_{\Sigma} $ on $\Sigma$ as the diagonal part of $U(1)_R \times 
SO(2)_{\Sigma}$, the adjoint matter $\phi$ has spin $r/2$ after the topological twist. In particular
the $R$-charge has to be $r=2$ for matching with the complex Chern-Simons theory where $\phi$ is a one form.
Thus the  twisted $\mathcal{N}=2$ Chern-Simons theory with an adjoint matter $\phi$ 
with $U(1)_R$ charge $r=2$ gives another description of the 2d TQFT that comes 
from the complex Chern-Simons theory. 
As remarked above, the complex Chern-Simons theory in the Hermitian description has
$SO(2)$ symmetry which rotates the one form components of $\phi$.  
In $\mathcal{N} =2$ Chern-Simons theory this symmetry is nothing but  
the $U(1)$ flavor symmetry of the adjoint chiral multiplet. 
As was proposed in \cite{Gukov:2015sna}, this $U(1)$ symmetry can be used to regularize the problem
of divergence due to the fact that the Hilbert space of complex Chern-Simons theory is infinite dimensional. 
Let us introduce the equivariant parameter $t := e^{-m}$ for the $U(1)$ rotation, 
where the parameter $m$ can be regarded as the mass for the adjoint matter\footnote{In the Nekrasov partition 
function of five dimensional SUSY Yang-Mills theory with the adjoint hypermultiplet 
a similar equivariant parameter appears.}.  The parameter $t$ is physically regarded as a Wilson 
loop of a background gauge field of $U(1)$ flavor symmetry. 
Then the corresponding 2d TQFT computes 
\beq
Z(\Sigma_g) = \trace_{\mathcal{H}} e^{- \beta H - m F} = \sum_{n=0}^\infty t^n \dim \mathcal{H}^{(n)},
\eeq
where $F$ is the charge of flavor symmetry and $\mathcal{H}^{(n)}$ is the charge $n$ sector of the physical Hilbert space. 
Since we have a smooth $t \to 0$ limit, which is the decoupling limit of the adjoint matter that gives the pure Chern-Simons 
theory, no negative powers of $t$ appear. The underlying algebra of this 2d TQFT is called
equivariant Verlinde algebra in \cite{Gukov:2015sna}. As we will see in our computation based on
Bethe/Gauge correspondence, the $U(1)$ equivariant parameter $t$ corresponds 
to the parameter of the Hall-Littlewood polynomial\footnote{See Appendix B for a definition and basic properties of $P_\lambda(x,t)$.}
$P_\lambda(x,t)$ where $t \to 0$ limit gives the Schur function $s_\lambda(x)$.

\subsection{Coulomb branch limit of superconformal index}

Now let us see the other side of 6d theory on $L(\kappa,1) \times \Sigma \times S^1$.
The superconformal index of four dimensional $\mathcal{N}=2$ theory is defined 
as the partition function on $S^3 \times S^1$, where we take the trace over $S^1$ direction
regarded as time coordinate. When the $\mathcal{N}=2$ superconformal theory is of class $\mathcal S$,
the superconformal indices give a 2d TQFT on the punctured Riemann surface $\Sigma_{g,n}$ associated with
the class $\mathcal{S}$ theory \cite{Gadde:2009kb, Benini:2011nc}. 
This is regarded as a TQFT version of AGT correspondence,
where conformal blocks are replaced by topological correlation functions. As a 2d TQFT 
the basic ingredients are the indices for the superconformal theories
coming from the genus zero surface with three punctures $\Sigma_{0,3}$,
which are identified with the topological three point functions $C_{\mu\nu\lambda}$.
The associativity condition for $C_{\mu\nu\lambda}$ is equivalent to the $S$-duality of
the class $\mathcal S$ theories. 
One can also consider the index on the lens space by
introducing the orbifold action on $S^3$ \cite{Alday:2013rs, Razamat:2013jxa}.
In general the superconformal index has three fugacities $\mathfrak{p}, \mathfrak{q}$ and $\mathfrak{t}$ 
\cite{Romelsberger:2005eg, Kinney:2005ej}.
There is a special limit called Coulomb branch limit which is defined by 
$\mathfrak{p}, \mathfrak{q},\mathfrak{t} \to 0$ while $t := \mathfrak{p}\mathfrak{q}/\mathfrak{t}$ fixed
\cite{Gadde:2011uv}.
According to the proposal in \cite{Gukov:2016lki} the $U(1)$ equivariant parameter $t$ is identified 
with the equivariant $U(1)$ parameter on the Chern-Simons side\footnote{In the theory of the superconformal index
there is a so-called Hall-Littlewood slice \cite{Gadde:2011uv}. Though the Hall-Littlewood polynomials are 
featured in the present paper, this has nothing to do with the Hall-Littlewood slice.}
The significant feature of the Coulomb branch limit is that the hypermultiplet does not contribute in the limit 
except the zero mode contributions. 
The superconformal theory obtained by twisted compactifications on $\Sigma_{0,3}$ of
6d $\mathcal{N}=(2,0)$ theory of type $A_{N-1}$ is called $T_N$ theory. 
When $N >2$ the theory does not allow the Lagrangian description and there is no
weak coupling region. In \cite{Gukov:2016lki} the computation of the superconformal indices for
$T_3$ theory has been made by invoking the Argyres-Seiberg duality that allows a weak coupling region. 
Unfortunately this approach cannot be generalized to $T_N$ theory for $3 < N$. 
In this paper we compute the partition function of $U(4)$ theory which is expected to 
match with the superconformal indices of $T_4$ theory.

\subsection{Organization of the paper}

This paper is organized as follows;
In the next section we review localization formula for $\mathcal{N} =2$ Chern-Simons matter theories in general. 
The final result involves an infinite magnetic sum of the multi-contour integrals which are called Jeffrey-Kirwan (JK) residues.
In section 3, we evaluate the integral of localization formula by the direct JK residue computation.
For technical reason,  the computation is possible for rank one case, namely $SU(2)$ theory.
We find a complete agreement with the result in \cite{Andersen:2016hoj} based on the geometry of $SU(2)$ Hitchin system. 
Section 4 is the main part of the paper; we use Bethe/Gauge correspondence to compute the structure
constants of the equivariant $U(N)$ Verlinde algebra. The localization formula shows
that the equivariant Verlinde algebra is realized by the algebra of Hall-Littlewood polynomials
$P_\lambda(x,t)$ with the specialization by the Bethe roots of the $q$-boson model, where $N$ corresponds to 
the number of excitations. 
Namely we substitute the solutions to the Bethe ansatz equation to the symmetric polynomial $P_\lambda(x,t)$.
After the specialization there arise relations among $P_\lambda(x,t)$ which are related by the affine Weyl group of $A_{N-1}$
acting on the partition $\lambda$ \cite{Korff:2013rsa}. 
We can understand these relations 
as a result of the quotient by the ideal $\mathcal{I}_{N, \kappa}$ determined by the space of 
Bethe roots. The characterization of the space by an ideal of the polynomial algebra
is one of the basic ideas in algebraic geometry. In fact this is a generalization of
what Gepner showed for the Verlinde algebra (fusion ring) \cite{Gepner:1990gr}, where the ideal is 
generated by derivatives of a potential $W(x)$ \cite{Intriligator:1991an, Witten:1993xi}. 
After taking the relation of $U(N)$ and $SU(N)$ theories into account, we can confirm the agreement of our result 
with the superconformal indices of $T_2$ and $T_3$ theories. 
In section 5, we discuss several aspects of the equivariant $U(N)$ Verlinde algebra, such as 
the recurrence relation among genus $g$ partition functions and the level-rank duality.
Finally backgrounds of 2d TQFT and the Hall-Littlewood polynomials are collected in appendices.


\section{Localization of topologically twisted Chern-Simons-matter theories  } 
In this section we consider topologically twisted Chern-Simons-matter (CS-matter) theories on $ \Sigma_g \times S^1$. 
The $R$-symmetry of $\mathcal{N}=2$ supersymmetric theory in 3 dimensions is $U(1)_R$ and the topological twist is 
made along $ \Sigma_g$ with local Lorentz symmetry $U(1)_{\rm spin}$\footnote{Topological twist on general three manifold is 
more non-trivial, since the local Lorentz symmetry is $SU(2)_{\rm spin}$.}. 
Namely we redefine the local Lorentz symmetry on $ \Sigma_g$ as the diagonal subgroup of $U(1)_R \times U(1)_{\rm spin}$. 
The observables of the CS-matter  theories are supersymmetric Wilson loops $W_{\lambda}$.
Here $\lambda$ expresses a representation of the gauge group $G$ and 
operators $\mathcal{O}_{\lambda}$'s in the correlation function are either supersymmetric Wilson loops or background flavor Wilson loops.  
Supersymmetric localization can be applied to correlation functions of $W_{\lambda}$ wrapping on $S^1$ 
and located at a point of $\Sigma_g$. 
When the rank of $G$ is $N$, the path integral reduces to $N$-dimensional contour integral (more precisely Jeffrey--Kirwan residue) 
by the localization formula  \cite{Benini:2015noa,  Benini:2016hjo, Closset:2016arn};
\begin{align}
\langle \prod_{i=1}^n \mathcal{O}_{\mu_i} \rangle_g 
&= \frac{1 }{|W(G)|}  \oint_{\mathrm{JK}(\eta)}  \prod_{a=1}^N \frac{d x_{a}}{2\pi i x_a} \sum_{\mathbf{k} \in \Gamma (G^{\vee})}
\left( \prod_{i=1}^n  \mathcal{O}_{\mu_i}(x, t) \right) e^{-S^{(\mathbf{k})}_{cl}}   \nonumber \\
&~~~~~~ \times Z^{(\mathbf{k})}_{\mathrm{vec}} (x, g) Z^{(\mathbf{k})}_{\mathrm{chi}} (x, t, g, r)     H (x, \kappa, t)^g,
\label{PF2d}
\end{align}
where $|W(G)|$ is the order of the Weyl group of $G$.
The integration variables $x_a$'s are saddle point values of  the Wilson loops associated to
 the $a$-th $U(1)$ Cartan  of $G$ and  
 the choice of the contour $\mathrm{JK}(\eta)$ is determined by an $N$-dimensional vector $\eta$.
The set of vectors
$\mathbf{k} \in \Gamma (G^{\vee})$ represents an element of the magnetic lattice of $G$ and 
the integrand comes from several multiplets in this susy model:
$Z^{(\mathbf{k})}_{\mathrm{vec}}$ is the one-loop determinant of the super Yang-Mills fields with 
the magnetic charge $\mathbf{k} =(k_1,\cdots, k_N)$:
\bel
Z^{(\mathbf{k})}_{\mathrm{vec}}(x, g)=
 (-1)^{\sum_{\alpha > 0} \alpha (\mathbf{k}) }    \prod_{\alpha \neq 0  } (1-x^{\alpha} )^{1-g}~,
\ee
where $\sum_{\alpha > 0} $ and $\prod_{\alpha \neq 0  }$  express summation over the positive root vectors and product over  the root vectors, respectively. 
 $x^{\alpha}$ stands for a paring of the Cartan part of Wilson loop $x$ and a root  $\alpha$.
$Z^{(\mathbf{k})}_{\mathrm{chi}}(x,t,g,r)$ is the one-loop determinant of a chiral multiplet in a representation $\mathbf{R}$ of the Lie algebra of $G$:
\bel
Z^{(\mathbf{k})}_{\mathrm{chi}}(x,t,g,r)=
\prod_{\rho \in \Delta (\mathbf{R}) }  \left( \frac{x^{\frac{\rho}{2}} t^{\frac{1}{2}}  }{1-  x^{\rho} t } \right)^{\rho (\mathbf{k}) +(1-g)(1-r)}~,
\ee
where $\Delta(\mathbf{R})$ expresses the set of weight vectors of the representation $\mathbf{R}$  and $r$ is $R$-charge for the lowest component scalar in the chiral multiplet.
 $x^{\rho}$ stands for the paring of the Cartan part of Wilson loop $x$  and a weight  $\rho$.
The parameter $t$ originates in the  background  flavor Wilson loop. 
In general, one can introduce background $U(1)$ flavor Wilson loops for the Cartan part of the flavor symmetry.
Later we consider the adjoint representation, which has only $U(1)$ flavor symmetry.  
The $Q$-closed action is a sum of the (mixed) Chern-Simons terms in three dimensions. 
Here $Q$ is a generator of supersymmetric transformation used in the localization computation.
Their saddle point values are written by $x_a$ and $t$
\begin{align}
e^{-S^{(\mathbf{k})}_{cl}}=
 \left( \prod_{a,b=1}^N x_a^{\kappa^{a b} k_b} \right)  t^{ \kappa_{(\mathrm{rf})} (g-1)} \cdots
\end{align}
where  
$\kappa^{a b}:=\kappa \mathrm{Tr}(H_a H_b)$ and $\{ H_{a} \}_{a=1}^N$ represents the Cartan  part of the Lie algebra of $G$ in the Chevalley basis. 
$\kappa$  and $\kappa_{(\text{rf})}$ are respectively gauge CS level and  
mixed CS level between flavor symmetry and $R$-symmetry ((rf)-mixed CS level). 
The symbol "$\cdots$" stands for other mixed CS terms which 
are not included in the model we will treat in the following sections.
We will also choose $\kappa_{(\mathrm{rf})} =\frac{N^2(1-r)}{2}$ for $G=U(N)$ and $\kappa_{(\mathrm{rf})} =\frac{(N^2-1)(1-r)}{2}$ for $G=SU(N)$
  when we look at the relation with the Coulomb branch limit of lens space index.  
But it is easy to recover genus $g$ partition function with generic value of $\kappa_{(\mathrm{rf})}$, 
because the (rf)-mixed CS term is independent of integration variables and magnetic charges.  
Finally $H(x, \kappa,t)$ is the Hessian of the effective twisted superpotential $W_{\text{eff}}(x)$ which comes from integration over gaugino zero modes;
\begin{align}
H(x, \kappa,t):=\det_{a, b} \left(  \frac{(2\pi i)^2 \partial^2 W_{\mathrm{eff}}}{\partial {\log x_a} \partial { \log x_b}} \right)=
\det_{a, b} \left( \kappa^{a b}  + \sum_{\alpha \in \Delta(\mathbf{R})} \rho^{a} \rho^{b} \frac{1}{2}
\left( \frac{1+ t x^{\rho} }{1- \, t x^{\rho}} \right)  \right) ~,
\end{align}
with
\begin{align}
(2 \pi i)^2 W_{\text{eff}}(x, \kappa, t)
=&
\frac{1}{2} \sum_{a,b=1}^N \kappa^{ab} (\log x_a)( \log x_b)  - 2 \pi^2 \sum_{\alpha >0} \alpha \nonumber \\
&~~~~~  +  \sum_{\rho \in \Delta(\mathbf{R})} \left( \mathrm{Li}_2 (x^{\rho} t)+\frac{1}{4} (\rho(\log x)+\log t)^2 \right) 
+\cdots ~.
\label{eq:twistSP}
\end{align}
In \eqref{eq:twistSP},  ellipse "$\cdots$" stands for the gauge flavor mixed CS-term which is taken as zero in our calculation.

If  the magnetic sum is performed before the evaluation of the integral and the contour
is deformed to enclose  saddle point configurations of the effective twisted superpotential $e^{2 \pi i\partial_{\log x_a} W_{\text{eff}}}=1$,  
the correlation function is expressed as
\begin{align}
\langle \prod_{i=1}^n \mathcal{O}_{\mu_i}   \rangle_{g} 
&= \frac{1 }{|W(G)|}  \sum_{x_* \in \mathrm{Sol} } \oint_{x=x_*}  \prod_{a=1}^N \frac{d x_{a}}{2\pi i x_a} \left( \prod_{i=1}^n \mathcal{O}_{\mu_i} (x,t) \right)
\nonumber \\
& \qquad \qquad \times\left( \prod_{a=1}^{N}\frac{1}{1-e^{2 \pi i\partial_{ \log x_a} W_{\text{eff}}}} \right)  
     e^{-S^{(0)}_{cl}}  Z^{(0)}_{\mathrm{vec}}  Z^{(0)}_{\mathrm{chi}}     H^{g}   
\label{PFresum1}
 \\
&=  \sum_{ x \in \mathrm{Sol} }
\left( \prod_{i=1}^n \mathcal{O}_{\mu_i}  \right)    e^{-S^{(0)}_{cl}}  Z^{(0)}_{\mathrm{vec}}  Z^{(0)}_{\mathrm{chi}}      H^{g-1}  ~.
\label{PFresum2}
\end{align}
When the gauge group is non-Abelian,
the summation $\sum_{x_* \in \mathrm{Sol} }$ is taken over the roots of the saddle point equation 
of the twisted superpotential $e^{2 \pi i\partial_{\log x_a} W_{\text{eff}}}=1$ 
except for $x^{\alpha}=1$ for any root $\alpha$. 
If  the roots $x$ with $x^{\alpha}=1$ are included in the residue operation, we find that  
the genus one partition function $\langle 1 \rangle_{g=1}$ from the expressions  \eqref{PFresum1} and  
\eqref{PFresum2}  does not reproduce the correct Witten index and also  
the  higher genus partition functions $\langle 1 \rangle_{g \ge 2}$  do not agree with the results predicted 
from the the Coulomb branch limit of lens space indices in our models. 
Thus we have to remove  the roots satisfying $x^{\alpha}=1$  and $\mathrm{Sol}$ is given by
\bel
\mathrm{Sol}:=\Bigl\{x= (x_1, \cdots, x_N) \Big| e^{2 \pi i \partial_{ \log x_a} W_{\mathrm{eff}}} =1, \, a=1,\cdots, N, \,   x^{\alpha}\neq 1 \, \, \text{for all the root} \, \, \alpha \Bigr\}\slash \sim
\nonumber \\
\ee
Here "$\sim$" means that we identify solutions which are equal up to the Weyl permutation. 
Since the theory is topologically twisted and does not depend on the metric on Riemann surfaces,
we expect the correlation functions satisfy the axiom of 2d TQFT or equivalently the set of observables $\mathcal{O}_\lambda$ 's
gives a finite dimensional commutative Frobenius algebra.
In Appendix A we summarize properties of 2d TQFT used in this paper. 
Especially, the definitions of the structure constant $C^{\lambda}_{\mu \nu}$, the metric $\eta_{\mu \nu}$ and 
the handle operator $(H \cdot C)_{\mu}^{\, \nu}$ are given by \eqref{eq:Fusion}, \eqref{eq:metric} and \eqref{eq:handle}, respectively. 
Note that we assume that the reduction of the Chern-Simons-matter theory 
to 2 dimensions gives 2d TQFT and compute the partition functions 
and correlation functions in higher genus from genus zero two point and
three point functions, for which we employ the localization formula. 
It is an interesting problem to check 
that the predictions based on 2d TQFT agree with the result 
of the direct computations of the localization formula\footnote{A 2d TQFT which reproduces the localization computation of
twisted CS-matter theory with an adjoint matter of $R$ charge $r=2$ 
is constructed in  \cite{Andersen:2016hoj} based on the moduli space of the Higgs bundle. }.

\section{Direct (residue) computations in $SU(2)$ case}

Let us apply the localization formula in the last section to 
$SU(2)$ CS-matter theory with an adjoint chiral multiplet. 
Since $SU(2)$ is rank one, the residue evaluation and the saddle point are relatively simple.
Unfortunately the direct computation in this section gets technically involved for higher rank gauge group.
We can evaluate the genus $g$ partition function by two methods: 
one is gluing the genus zero three point functions and the other is the direct residue evaluation of the 
higher genus  partiton function in the summed form \eqref{PFresum1}. 
Each method has its advantages and disadvantages.
In the first method we make use of the properties of 2d TQFT and once we obtain
the genus zero three point functions it is rather easy to compute the partition functions
for any higher genus. However, the computation becomes quite involved for higher level $\kappa$,
since the dimensions of the Frobenius algebra $\mathcal{A}$ increase  with $\kappa$.
On the other hand, in the second method 
we do not have to rely on 2d TQFT structure and there is no complication with higher $\kappa$
mentioned above. But the higher genus computations are difficult in this case.
Thus we can obtain the result for arbitrary level $\kappa$ but only for lower genera.

The saddle point equation of  the twisted superpotential is given in the $SU(2)$ model
\begin{align}
\exp \left( 2\pi i \frac{\partial W_{\mathrm{eff}} }{\partial \log x} \right)=x^{2\kappa+4} \left(\frac{1-t x^{-2}}{1-t x^2} \right)^2=1.
\label{eq:saddleSU2}
\end{align}
First, we  shall directly evaluate  the residue  in the resumed form \eqref{PFresum1}.   
We can write down the genus $g$ partition function of this $SU(2)_{\kappa}$ theory with $R$-charge $r=2$  
\begin{align}
\langle \prod_{i=1}^n \mathcal{O}_{\mu_i}  \rangle_{g} 
&=  \sum_{x_* \in \mathrm{Sol}} \oint_{x= x_* } \frac{dx}{2\pi i }
 \left( \prod_{i=1}^n \mathcal{O}_{\mu_i} (x,t) \right)  \omega_g (x,t,\kappa)\,,
\end{align}
where
\begin{align} 
\omega_g (x, t, \kappa):=   
\frac{1 }{2x}   
   \left({1- x^{2\kappa+4} \left(\frac{1-t x^{-2}}{1-t x^2}  \right)^2} \right)^{-1}
   \left[ (1-t)  \prod_{d= \pm1}  (1-x^{2d}) (1-t   x^{2d} ) \right]^{1-g} 
  H (x, \kappa, t)^g\,,  \nonumber \\ 
\end{align}
with
\begin{align}
H (x, \kappa ,t)=2 \kappa+2 \frac{1+ t x^2}{1-t x^2}+2 \frac{1+ t x^{-2} }{1-{t} x^{-2}}.
\end{align}
The roots of the effective twisted superpotential except $x^{\alpha}=1$ are collected into 
the set "Sol"
\bel
{\mathrm{Sol}}=\Bigl\{ x \Big|x^{2 \kappa+4}-2 t x^{2 \kappa+2}+ t^2 x^{2 \kappa}-t^2 x^4+2 t x^2-1=0, \, x^2 \neq 1  \Bigr\}~.
\label{rootSU2c}
\ee
Then we find that $|\mathrm{Sol}|/|W(SU(2))|=\kappa+1$ reproduces the correct Witten index  for the $SU(2)$ theory.
The higher genus partition function is given by 
\bel
Z^{SU(2)_{\kappa}}_{g}= \sum_{x_* \in \mathrm{Sol}} \oint_{x= x_* } \, \omega_g (x,t,\kappa)~.
\ee
Let us evaluate these higher genus partition functions.  
When $g \ge 2$, the poles of $\omega_g (x,t,\kappa)$ are located at 
$\{\pm1, \pm {t}^{1/2}, \pm {t}^{-1/2} \} \cup {\mathrm{Sol}}$ on the Riemann sphere $\C \cup \{ \infty \} \ni x$ 
and we have 
\begin{align}
Z^{SU(2)_{\kappa}}_g=\sum_{x_* \in \mathrm{Sol}} \oint_{x= x_* } \frac{dx}{2\pi i} \,
\omega_g(x,t,\kappa)= - \sum_{x_*=\pm t^{\frac{1}{2}}, \pm t^{-\frac{1}{2}}, \pm 1 \, }  \oint_{x= x_* }  \frac{dx}{2\pi i} \, \omega_g(x,t,\kappa), \quad (g \ge 2)\,.
\end{align}
For example,  we can show the partition functions with $g=2,3$ explicitly 
\begin{align}
Z^{SU(2)_{\kappa}}_{g=2}&=\frac{1}{6 (t-1)^6 (t+1)^3}
\Bigl[ \kappa ^3 \left(1-t^2\right)^3+6 \kappa ^2 \left(t^2-1\right)^2 \left(t^2+1\right)-
 \kappa  (11 t^6-36 t^5 \nonumber \\ 
&
-9   t^4+9 t^2+36 t-11)
+6 \left(-16 t^{\kappa +3}+t^6-6 t^5+15 t^4-4 t^3+15 t^2-6 t+1\right) \Bigr]\,,
\label{eq:generalSU22}
\end{align}
and 
\begin{align}
Z^{SU(2)_{\kappa}}_{g=3}&=
\frac{1}{180 (t-1)^{12} (t+1)^6} \Bigl[
 \kappa ^6 \left(t^2-1\right)^6-12 \kappa ^5 (t-1)^5 (t+1)^7 \nonumber \\
&+10 \kappa ^4 (t-1)^4 \left(7 t^2-2 t+7\right) (t+1)^6 -240 \kappa ^3 \left(t^2-1\right)^3 \left(t^6-t^5-4 t^4-10 t^3-4 t^2-t+1\right) \nonumber \\
&+\kappa^2 \left(t^2-1\right)^2 (31680 t^{\kappa +4}+469 t^8-2280 t^7+44 t^6-6360 t^5+7614 t^4 \nonumber \\
&-6360 t^3+44 t^2-2280 t+469)  
-36 \kappa    \left(t^2-1\right) (4160 t^{\kappa +4}+3200 t^{\kappa +5}+4160 t^{\kappa +6}+13 t^{10} \nonumber \\
&-114 t^9+361 t^8+296 t^7+2986 t^6 
+1556 t^5+2986 t^4+296 t^3+361 t^2-114
   t+13) \nonumber \\
&+180 ( 960 t^{\kappa +4}+1536 t^{\kappa +5}+2944 t^{\kappa +6}+1536 t^{\kappa +7}+960 t^{\kappa +8}+64 t^{2 \kappa +6}+t^{12}-12 t^{11} \nonumber \\
&+66 t^{10}-220 t^9-465   t^8-2328 t^7-2084 t^6-2328 t^5-465 t^4-220 t^3+66 t^2-12 t+1) \Bigr] .
\label{eq:generalSU23}
\end{align}
Eqs.\eqref{eq:generalSU22} and \eqref{eq:generalSU23} are consistent with the results in \cite{Andersen:2016hoj},
which were obtained from the geometry of the moduli space of $SU(2)$ Hitchin system.

Next we calculate  partition functions based on 2d TQFT structure, namely by gluing genus zero three point functions 
obtained by evaluating the Jeffrey-Kirwan residues. As we remarked at the beginning of the section,
the computations are made with level by level, since the underlying Frobenius algebra $\mathcal{A}$ depends on the level.
The localization formula \eqref{PF2d} for the $SU(2)_{\kappa}$ model with $r=2$ tells us that 
genus zero three point function is
\begin{align}
\langle \mathcal{O}_{\mu}  \mathcal{O}_{\nu} \mathcal{O}_{\lambda} \rangle_{g=0} 
&= \frac{1-t}{2}  \oint_{\mathrm{JK}(\eta)} \frac{dx}{2\pi i x}~
\mathcal{O}_{\mu}(x,t)  \mathcal{O}_{\nu}(x,t)  \mathcal{O}_{\lambda}(x,t) \cdot (1-x^2) (1-x^{-2}) \nonumber \\
& \times  \sum_{ k \in \Z}   x^{2 \kappa k }  
 \left( \frac{  x  }{1-t \,  x^2 } \right)^{ 2 k -1} 
 \left( \frac{  x^{-1}  }{1-t \,  x^{-2} } \right)^{- 2 k -1}    \,.
\label{eq:SU2corre}
\end{align}
When we choose a vector $\eta<0$, the Jeffrey-Kirwan residue operation  is evaluated  at the poles $ x=\pm {t}^{\frac{1}{2}}, 0$.
On the other hand, when we choose a vector $\eta>0$,  the residue is evaluated at $x=\pm {t}^{-\frac{1}{2}}, \infty$. 
Since there are no poles except for $x=\pm {t}^{\frac{1}{2}}, 0, \pm {t}^{-\frac{1}{2}}, \infty$ in 
the genus zero case \eqref{eq:SU2corre},
the genus zero correlation functions evaluated at positive and negative $\eta$  cause the same result 
up to an overall sign. In the next section we will compute $U(2)$ case. 
The comparison of the following results with those in the next section 
gives a supporting evidence for the relation of $SU(N)$ and $U(N)$ partition functions 
derived in the next section (see \eqref{UNSUNnorm}). 
In fact the (mutually distinct) roots $y_i$ of the characteristic polynomial of the handle operator $(H \cdot C)$ are
related by
\beq
\left( \frac{\kappa}{2} \right) y_i^{SU(2)} = (1-t) \cdot y_i^{U(2)} 
\eeq
for $\kappa = 2,3,4$. 
Note that the dimensions of the Frobenius algebra $\mathcal{A}$ are different for $SU(2)$ and $U(2)$
and the multiplicity of each root $y_i$ is also different. 
The multiplicity of $U(N)$ theory is $\kappa/N$ times that of $SU(N)$ theory.

\subsubsection*{\underline{Level $\kappa=2$}}
First we study $\kappa=2$ case.  
The field configuration of a supersymmetric Wilson loop $W_{\lambda}=\mathrm{Tr}_{\lambda} x$  in a representation $\lambda$  
is symmetric under the exchange $x \leftrightarrow x^{-1}$ at the saddle points and 
the Wilson loop algebra consists of  functions of $x+x^{-1}$. 
We can also  include the background Wilson loop $t$ for the $U(1)$ flavor symmetry in the 
correlation function. Thus an
 element of the Wilson loop algebra takes value in $\C[[t]][x, x^{-1}]^{\mathfrak{S}_2}$. The equivalence relation $\mathcal{I}$ for this theory can be constructed by 
the saddle point equation  \eqref{eq:saddleSU2}
\begin{align}
\left(x+\frac{1}{x} \right) \left(x^2 +\frac{1}{x^2}-2t\right)=0~.
\label{eq:equiv1level2}
\end{align}
Here we have  removed  $(x^2-1)$ to produce the ideal ${\cal I}$ correctly 
and the algebra of the Wilson loops is given by
\begin{align}
\mathcal{A}=\mathbb{C}[[t]][x, x^{-1}]^{\mathfrak{S}_2}/ \langle (x+{x}^{-1} ) (x^2 +x^{-2}-2t ) \rangle \,.
\label{eq:basisSU2k2}
\end{align}
We can take  a basis of \eqref{eq:basisSU2k2} as $ \{ 1,  x+x^{-1},  x^2+x^{-2} \} $.     
Then the number of generators  is equal to the genus one partition function (Witten index) 
$Z_{g=1}=3$ for $\kappa=2$.
Products among  $ \{ 1,  x+x^{-1},  x^2+x^{-2} \} $ lead to  structure constants; 
\begin{align}
&C^{1}_{\mu \nu}=\left(
\begin{array}{ccc}
 1 & 0 & 0 \\
 0 & 2 & 0 \\
 0 & 0 & 4t  \\
\end{array}
\right), \, 
&C^{2}_{\mu \nu}=\left(
\begin{array}{ccc}
 0 & 1 & 0 \\
 1 & 0 & 2t \\
 0 & 2t & 0  \\
\end{array}
\right), \quad
 &C^{3}_{\mu \nu}=\left(
\begin{array}{ccc}
 0 & 0 & 1 \\
 0 & 1 & 0 \\
 1 & 0 & 2(t-1)  \\
\end{array}
\right).
\label{eq:fusionSU2k2}
\end{align}
From \eqref{eq:SU2corre} with insertion of operators $ \{ 1,  x+x^{-1},  x^2+x^{-2} \} $, we obtain the metric
\begin{align}
&\eta_{\mu \nu }=\left(
\begin{array}{ccc}
 1-t^4 & 0 & -t^5+3 t^4+2 t^3-2 t^2-t-1 \\
 0 & (1-t)^3 (t+1)^2 & 0 \\
 -t^5+3 t^4+2 t^3-2 t^2-t-1 & 0 & 2 \left(1  +2 t+t^2-4 t^3 -t^4+2 t^5-t^6\right) \\
\end{array}
\right). 
\label{eq:metricSU2k2}
\end{align}
From \eqref{eq:fusionSU2k2} and  \eqref{eq:metricSU2k2}, we 
can compute the characteristic polynomial $\det(y I-H \cdot C)=(y-y_1)^2(y-y_2)$ with
\begin{align}
\quad y_1=\frac{4}{(1-t)^3 (t+1)}
, \quad y_2=\frac{2}{(1-t) (t+1)^3}\,,
\end{align}
and  the partition function with genus $g$  is given by
\begin{align}
Z_{g}= 2 y_1^{g-1}+y_2^{g-1} \,.
\end{align}
For example, we shall show partition functions in lower genera
\begin{align}
&Z_{g=0}= 1-t^4 \,,\,\,
Z_{g=1}=3 \,,\,\,
Z_{g=2}= \frac{2 (5 + 6 t + 5 t^2)}{ (1 - t^2)^3} \,,\\
&Z_{g=3}= \frac{4 \left(9 t^4+28 t^3+54 t^2+28 t+9\right)}{ (1-t^2)^6} \,,\\
&Z_{g=4}=\frac{8 \left(17 t^{6}+90 t^{5}+255 t^4+300 t^3+255 t^2+90 t+17\right)}{\left(1-t^2\right)^9}\,.
\end{align}
This result  reproduces\footnote{It seems there are a few typos in the table.}
 table 1 in \cite{Gukov:2016lki} and Eqs.\eqref{eq:generalSU22} \eqref{eq:generalSU23}.

In a similar manner, we can evaluate the genus $g$ partition functions for the 
$SU(2)$ models with  $\kappa=3, 4$. 
We summarize our results in these models: the algebra of Wilson loops ${\cal A}$,  
structure constants $C_{\mu\nu}^{\lambda}$ in a basis $\{1, x^{l}+x^{-l} \}_{l=1, \dots, \kappa}$, 
the metric $\eta_{\mu\nu}$ and the characteristic polynomial of the handle operator.

\subsubsection*{\underline{Level $\kappa=3$}}
\begin{itemize}
\item The algebra of Wilson loops 
\begin{align}
\mathcal{A}=\C[[t]][x, x^{-1}]^{\mathfrak{S}_2}/ \langle (1-t)^2   + (1- 2 t) ( x^2+x^{-2}) + x^4+x^{-4} \rangle\,.
\label{eq:basisSU2k3}
\end{align}
\item The structure constants $C^{\lambda}_{\mu \nu}$ in a basis  $ \{ 1,  x+x^{-1},  x^2+x^{-2}, x^3+x^{-3} \} $ 
{\footnotesize
\begin{align}
&C^{1}_{\mu \nu}=\left(
\begin{array}{cccc}
 1 & 0 & 0 & 0 \\
 0 & 2 & 0 & -(t-1)^2 \\
 0 & 0 & -t^2+2 t+1 & 0 \\
 0 & -(t-1)^2 & 0 & -2 t^3+5 t^2+1 \\
\end{array}
\right), \, 
C^{2}_{\mu \nu}=\left(
\begin{array}{cccc}
 0 & 1 & 0 & 0 \\
 1 & 0 & 1 & 0 \\
 0 & 1 & 0 & -t^2+4 t-1 \\
 0 & 0 & -t^2+4 t-1 & 0 \\
\end{array}
\right), \nonumber \\
 &C^{3}_{\mu \nu}=\left(
\begin{array}{cccc}
 0 & 0 & 1 & 0 \\
 0 & 1 & 0 & 2 t \\
 1 & 0 & 2 t-1 & 0 \\
 0 & 2 t & 0 & (t-1) (3 t+1) \\
\end{array}
\right), \quad 
C^{4}_{\mu \nu}=\left(
\begin{array}{cccc}
 0 & 0 & 0 & 1 \\
 0 & 0 & 1 & 0 \\
 0 & 1 & 0 & 2 (t-1) \\
 1 & 0 & 2 (t-1) & 0 \\
\end{array}
\right).
\end{align}
}
\item The metric
{\tiny
\begin{align}
&\eta_{\mu \nu }=\left(
\begin{array}{cccc}
 (1-t) \left(t^2+t+1\right) & 0 & (1-t) \left(t^3-t^2-2 t-1\right) & 0 \\
 0 & (1-t) \left(t^3+t^2+1\right) & 0 & (1-t) \left(t^4-t^3-4 t^2-t-1\right) \\
 (1-t) \left(t^3-t^2-2 t-1\right) & 0 & (1-t) \left(t^4-2 t^3-t^2+3 t+2\right) & 0 \\
 0 & (1-t) \left(t^4-t^3-4 t^2-t-1\right) & 0 & (1-t) \left(t^5-2 t^4-2 t^3+8 t^2+5 t+2\right) \\
\end{array}
\right).
\end{align}
}
\item The characteristic polynomial 
\begin{align} 
\det(y I-H \cdot C)=(y-y_+)^2(y-y_-)^2\,,
\end{align}
with
\begin{align} 
y_{\pm}=\frac{4 t^3+9 t^2+9 t+5 \pm \left(3 t^2+5 t+1\right)\sqrt{4 t+5}}{\left(1-t^2\right)^3}\,.  \label{SU2level3}
\end{align}
\end{itemize}

\subsubsection*{\underline{Level $\kappa=4$}}
\begin{itemize}
\item  The algebra of Wilson loops
\begin{align}
\mathcal{A}=\C[[t]][x, x^{-1}]^{\mathfrak{S}_2}/ \langle (1-t)^2\left(x+x^{-1} \right)+(1-2t)\left(x^3+x^{-3}\right)+x^5+x^{-5} \rangle\,.
\label{eq:basisSU2k4}
\end{align}
\item 
The structure constants $C^{\lambda}_{\mu \nu}$ in a basis  $ \{ 1,  x+x^{-1},  x^2+x^{-2}, x^3+x^{-3}, x^4+x^{-4} \} $
{\tiny
\begin{align}
&C^{1}_{\mu \nu}=\left(
\begin{array}{ccccc}
 1 & 0 & 0 & 0 & 0 \\
 0 & 2 & 0 & 0 & 0 \\
 0 & 0 & 2 & 0 & -2 (t-1)^2 \\
 0 & 0 & 0 & -2 (t-2) t & 0 \\
 0 & 0 & -2 (t-1)^2 & 0 & -4 t^3+10 t^2-4 t+2 \\
\end{array}
\right),  
&C^{2}_{\mu \nu}=\left(
\begin{array}{ccccc}
 0 & 1 & 0 & 0 & 0 \\
 1 & 0 & 1 & 0 & -(t-1)^2 \\
 0 & 1 & 0 & (2-t) t & 0 \\
 0 & 0 & (2-t) t & 0 & -2 (t-2) t^2 \\
 0 & -(t-1)^2 & 0 & -2 (t-2) t^2 & 0 \\
\end{array}
\right), \nonumber 
\end{align}
\begin{align}
 &C^{3}_{\mu \nu}=\left(
\begin{array}{ccccc}
 0 & 0 & 1 & 0 & 0 \\
 0 & 1 & 0 & 1 & 0 \\
 1 & 0 & 0 & 0 & -t^2+4 t-1 \\
 0 & 1 & 0 & -t^2+4 t-2 & 0 \\
 0 & 0 & -t^2+4 t-1 & 0 & -2 (t-3) (t-1) t \\
\end{array}
\right), 
 &C^{4}_{\mu \nu}=\left(
\begin{array}{ccccc}
 0 & 0 & 0 & 1 & 0 \\
 0 & 0 & 1 & 0 & 2 t \\
 0 & 1 & 0 & 2 t-1 & 0 \\
 1 & 0 & 2 t-1 & 0 & (t-1) (3 t+1) \\
 0 & 2 t & 0 & (t-1) (3 t+1) & 0 \\
\end{array}
\right), \nonumber 
\end{align}
\begin{align}
&C^{5}_{\mu \nu}=\left(
\begin{array}{ccccc}
 0 & 0 & 0 & 0 & 1 \\
 0 & 0 & 0 & 1 & 0 \\
 0 & 0 & 1 & 0 & 2 (t-1) \\
 0 & 1 & 0 & 2 (t-1) & 0 \\
 1 & 0 & 2 (t-1) & 0 & (t-1) (3 t-1) \\
\end{array}
\right)\,,
\end{align}
}
\item The metric 
{\tiny
\begin{align}
&\eta_{\mu \nu }=\left(
\begin{array}{ccccc}
 1-t^3 & 0 & t^3+t^2-t-1 & 0 & -t^4+t^3-t^2+t \\
 0 & -t^3+t^2-t+1 & 0 & -t^4+2 t^3-1 & 0 \\
 t^3+t^2-t-1 & 0 & -t^4-t^3-t^2+t+2 & 0 & -t^5+3 t^4+2 t^3-2 t^2-t-1 \\
 0 & -t^4+2 t^3-1 & 0 & -t^5+3 t^4-t^3-3 t^2+2 & 0 \\
 -t^4+t^3-t^2+t & 0 & -t^5+3 t^4+2 t^3-2 t^2-t-1 & 0 & -t^6+3 t^5-10 t^3+3 t^2+3 t+2 \\
\end{array}
\right)\,,
\end{align}
}

\item
The characteristic polynomial 
\begin{align} 
&\det(y I-H \cdot C)=(y-y_1)^2(y-y_2)^2(y-y_3) \,,\nonumber \\
&{\rm with}\quad y_{1}=\frac{4}{1-t^2}, \quad y_2=\frac{4 (t+3)}{(1-t)^3}, \quad y_3=\frac{t+3}{(1-t) (t+1)^3}\,.  \label{SU2level4}
\end{align}
\end{itemize}


\section{Equivariant $U(N)$ Verlinde algebra via Bethe ansatz}

When the rank $N$ of the gauge group $G$ is greater than one, the evaluation of the 
residue integral becomes difficult.
It is desirable to have alternative method to compute the correlation functions
and it is here that the Bethe/Gauge correspondence saves the day. 
In this section, we will evaluate partition functions of $U(N)_{\kappa}$ Chern-Simons theory with an adjoint chiral multiplet with  $r=2$ for $N=2, 3$ and $4$ with lower $\kappa$.
As we explain in Appendix A, the partition function of 2d TQFT is characterized by the structure constant $C^{\lambda}_{\mu \nu}$ and the metric $\eta_{\mu \nu}$. 
The Bethe/Gauge correspondence allows us to obtain these quantities from the algebra of Hall-Littlewood polynomials with the specialization 
on the set of explicit solutions (Bethe roots) to the Bethe ansatz equation.
A crucial fact is that in this approach we do not have to solve  the Bethe ansatz equation explicitly.
What we need is the generating relations among Hall-Littlewood polynomials with the specialization, which  mathematically define an ideal of 
the algebra of Hall-Littlewood polynomials. 
More precisely speaking the realization by the Hall-Littlewood polynomials is obtained,
when the adjoint matter $\phi$ has the $R$-charge $r=0$, while the equivariant Verlinde algebra
is related to the case $r=2$ \cite{Gukov:2015sna}. However, we can control the dependence of the Frobenius 
algebra structure on the $R$-charge of $\phi$, since in the localization formula the $R$-charge $r$ only appears in the power of
the difference product $\Delta(x,t)$ which can be expanded by the Hall-Littlewood polynomials.
It turns out that the structure constants of the algebra are universal in the sense that they are independent of $r$ and
the $r$  dependence appears in the metric (topological two point function).
Note that the three point functions also depend on $r$, since it is obtained by contracting the structure 
constants with the metric. 
Unfortunately it is difficult to compute 
the partition function of the $SU(N)$ theory with $N>2$ directly. 
But we can compare our results with those obtained by other methods,  after computing the genus $g$ partition function,
by using the relation of the $U(N)$ and $SU(N)$ partition functions, which we explain shortly below.

Let us propose the relation between $U(N)$ and $SU(N)$ partition functions on $  \Sigma_g \times S^1$.
We decompose  the Cartan part of  $U(N)$ Wilson loop $(x_1, \cdots, x_N)$ to central $U(1)$ Wilson loop $y$  and the Cartan part of $SU(N)$ Wilson loop $(\tilde{x}_1, \cdots, \tilde{x}_{N-1} )$ as
\begin{align}
x_1=  y  \tilde{x}_1, \quad  x_2=y \tilde{x}_1^{-1}  \tilde{x}_2, \, \, \cdots, \, \, x_{N-1}=y \tilde{x}_{N-2}^{-1}  \tilde{x}_{N-1}, \quad x_N=y  \tilde{x}_{N-1} 
\label{eq:decompo}\,.
\end{align}
Since the one-loop determinants and $H$ do not depend on $y$, the center $U(1)$ CS term only depends on $y$. 
The integration of $y$ leads to an  $SU(N)$ condition for the magnetic charge. 
There is also a  relation between Hessians of  $U(N)_{\kappa}$ and $SU(N)_{\kappa}$ CS-matter theories: 
\begin{align}
\det_{ab} \left( \frac{(2\pi i)^2\partial^2 W^{U(N)}_{\text{eff}}}{\partial \log x_a \partial \log x_b} \right)
=\frac{\kappa}{N} 
 \det_{ab} \left( \frac{(2\pi i)^2 \partial^2 W^{SU(N)}_{\text{eff}}}{\partial \log \tilde{x}_a \partial \log \tilde{x}_b} \right)\,,
\end{align}
and we connect  $SU(N)$ partition functions with $U(N)$ partition functions.
Then we obtain the following relation between   $SU(N)$ and  $U(N)$ partition functions
\bel
Z^{U(N)_{\kappa}}_g= \left( \frac{\kappa}{N} \right)^g (1-t)^{(g-1)(1-r) } Z^{SU(N)_{\kappa}}_g
\label{UNSUNnorm}\,.
\ee

By applying the resumed expression \eqref{PFresum2} to
the $U(N)_{\kappa}$ Chern-Simons theory 
with an adjoint chiral multiplet with integer R-charge $r$, 
the genus $g$ correlation functions are 
 given as a sum over the saddle points  of the twisted superpotential  $W_{\mathrm{eff}}$;
\begin{align}
\langle \prod_{i=1}^l \mathcal{O}_{\mu_i}(x,t) \rangle^{r}_{g}=\sum_{x \in \mathrm{Sol}}  \prod_{i=1}^l \mathcal{O}_{\mu_i}(x,t) 
\left( \prod_{a \neq b}^N (1-x_a x_b^{-1}) \prod_{a, b=1}^N (1-t x_a x_b^{-1})^{r-1} H^{-1}(x,t) \right)^{1-g} ,
\label{eq:corCSadj}
\end{align}
where $\text{Sol}$ is given as 
the set of the roots of the following saddle point equation 
\begin{align}
\exp \left(2\pi i \frac{\partial W_{\mathrm{eff}}}{\partial \log x_a} \right)= x^{\kappa}_a \prod_{b =1 \atop b \neq a}^N  \frac{ x_a-t x_b}{t x_a-x_b}=1, \quad a=1, \cdots, N.
\label{eq:betheqbosn}
\end{align}
As pointed out in \cite{Okuda:2013fea}, \eqref{eq:betheqbosn} coincides with the 
Bethe ansatz equation of the $q$-boson model with periodic boundary condition.
The $q$-boson model is a one-dimensional quantum integrable lattice model which is regarded as a non-linear deformation of the harmonic oscillator. 
Especially $U(1)$ flavor Wilson loop $t$ corresponds to the $q$-deformation parameter $q$ by $t=q^2$.
Parameters $N$ and $\kappa$ correspond to the particle number and the number of lattice sites. 
Although the Bethe ansatz equation \eqref{eq:betheqbosn} cannot be solved explicitly, we will show that three point functions for $r=2$
 are explicitly  calculable.  
We summarize important properties  of the $q$-boson model studied in \cite{Korff:2013rsa} to evaluate the partition functions. 
Let us introduce  $\mathcal{P}_{ N, \kappa}$ and $\widetilde{\mathcal{P}}_{ N, \kappa}$  as 
collections of non-negative integers $(\lambda^{1}, \lambda^{2}, \cdots, \lambda^{N}) $
\begin{align}
&\mathcal{P}_{N, \kappa}:=\{ \lambda=(\lambda^{1}, \lambda^{2}, \cdots, \lambda^{N}) |  \kappa \ge \lambda^{1} \ge \lambda^{2} \ge \cdots \lambda^{N} \ge 1 \}, \\
&\widetilde{\mathcal{P}}_{N, \kappa}:=\{ \lambda=(\lambda^{1}, \lambda^{2}, \cdots, \lambda^{N}) |  \kappa > \lambda^{1} \ge \lambda^{2} \ge \cdots \lambda^{N} \ge 0 \}.
\end{align}
A bijection $\tilde{} : \mathcal{P}_{N, \kappa} \to \widetilde{\mathcal{P}}_{N, \kappa}  $ which sends $\lambda \to \tilde{\lambda}$ is defined by eliminating  all the integer $\kappa$ from the partition $\lambda \in {\mathcal{P}}_{N, \kappa}$. 
Another  bijection  $ {}^* :  \mathcal{P}_{N, \kappa} \to \mathcal{P}_{N, \kappa} $ called $*$-involution is defined as some kind of an inverse operation of  the bijection $\mbox{}\tilde{}$\,, namely, 
$\lambda^*$ for $\lambda=(\lambda^1, \cdots, \lambda^N)$ is defined by 
the inverse image  of $(\kappa-\lambda^N, \cdots,\kappa- \lambda^1)  \in \widetilde{ \mathcal{P}}_{N, \kappa} $ by $ \tilde{}$.
From proposition 7.7 in \cite{Korff:2013rsa}, a basis of the algebra of the Wilson loops  can be taken as a set of Hall-Littlewood polynomials  $\{ P_{\lambda}(x,t) \}_{\lambda \in \mathcal{P}_{N, \kappa}}$, which means 
 the number of roots of \eqref{eq:betheqbosn} equals to the order of the set $\mathcal{P}_{N, \kappa}$. 
Especially, this means the genus one partition function is given by the number of elements of $\mathcal{P}_{N, \kappa}$;
\begin{align}
Z^{U(N)_{\kappa}}_{g=1}=\frac{(N+\kappa-1)!}{N! (\kappa-1)!}\,,
\end{align}
and the genus one partition function of the $SU(N)_{\kappa}$  theory
is written down by using the relation \eqref{UNSUNnorm} 
\begin{align}
Z^{SU(N)_{\kappa}}_{g=1}=\frac{(N+\kappa-1)!}{(N-1)! \kappa!}
\label{eq:WittenindexSU}\,.
\end{align}
This result  \eqref{eq:WittenindexSU}  correctly reproduces  Witten index  of  
  $\mathcal{N}=2$ $SU(N)_{\kappa}$ CS-matter theory.
Note that we have $Z^{SU(N)_{\kappa}}_{g=1} = Z^{U(\kappa)_{N}}_{g=1}$,
where the correspondence of the states is given by the transpose of the Young diagrams.

The structure constants $C^{\lambda}_{\mu \nu}(t)$ in this basis are defined by the expansion of products of Hall-Littlewood polynomials  
\begin{align}
 P_{\mu}(x,t) P_{\nu}(x,t) \equiv \sum_{\lambda \in \mathcal{P}_{N, \kappa} }  C^{\lambda}_{\mu \nu} (t)  P_{\lambda}(x,t), \quad \mu, \nu \in \mathcal{P}_{N, \kappa} \quad 
\mathrm{and} \quad x \in \mathrm{Sol}.
\label{eq:deffusion}
\end{align}
We use ``$\equiv$'' to emphasize equality up to the Bethe ansatz equation \eqref{eq:betheqbosn}.
An important property of $C^{\lambda}_{\mu \nu} (t)$  is that there exists $S_{\mu \nu}(t)$ which simultaneously diagonalizes the structure constants
\begin{align}
C^{\lambda}_{\mu \nu} (t)= \sum_{\sigma \in \mathcal{P}_{N, \kappa} } \frac{S_{\mu \sigma}(t) S_{\nu \sigma}(t) S^{-1}_{ \sigma \lambda}(t) } {S_{\emptyset \sigma}(t)}\,.
\label{eq:defVeralg}
\end{align}
Here $\emptyset:=(\kappa,\cdots,\kappa)$. 
Note that \eqref{eq:defVeralg}  is independent of $R$-charge.  
Then the associativity condition \eqref{eq:associative} immediately follows from \eqref{eq:defVeralg}.
It is also shown in \cite{Korff:2013rsa}  that   
\begin{align}
\sum_{x \in \mathrm{Sol} } \frac{P_{\lambda}(x,t) P_{\mu}(x,t) P_{\nu}(x,t)}{\langle \psi_N (x) |  \psi_N (x) \rangle}
=\frac{C^{\lambda^*}_{\mu \nu}(t)}{b_{\lambda}(t)},  \quad \lambda, \mu, \nu \in \mathcal{P}_{N, \kappa} \,.
\label{eq:three2}
\end{align}
Here $|  \psi_N (x) \rangle$ and $ \langle  \psi_N (x) |$ are respectively   
on-shell Bethe vector of $N$  particles  and dual Bethe vector  in the  $q$-boson model. 
$\langle \psi_N (x) |  \psi_N (x) \rangle$ is the inner product of these  two vectors and
$b_{\lambda}(t)$ is defined by
\begin{align}
&b_{\lambda}(t):=\prod_{i \ge 1} \prod_{j=1}^{m_i(\lambda)} (1-t^{j}), \quad m_i(\lambda):=\# \{l | \lambda_l=i  \}
\,.
\label{bmu}
\end{align}

We can relate genus zero three point functions with $r=0$  to the metric or genus zero two point functions with $r=2$ as follows.
The correlation functions \eqref{eq:corCSadj} with $r=0$ agree with the correlation functions of  the 
 $U(N)/U(N)$ gauged WZW-matter model with level $\kappa$ on genus $g$ Riemann surface introduced in \cite{Okuda:2013fea}. 
Then, it was shown in \cite{Okuda:2013fea} that the genus zero correlation functions with $r=0$ are expressed  as
\begin{align}
\langle \prod_i \mathcal{O}_{\mu_i} \rangle^{r=0}_{g=0}
=\sum_{x \in \mathrm{Sol} } \frac{\prod_i \mathcal{O}_{\mu_i}(x,t)}{\langle \psi_N (x) |  \psi_N (x) \rangle}\,.
\label{eq:three1}
\end{align}
From \eqref{eq:three2} and \eqref{eq:three1},  the  genus zero three point functions of 
$\{ P_{\lambda}(x,t) \}_{\lambda \in \mathcal{P}_{N, \kappa}}$ are 
given by
\begin{align}
\langle P_{\lambda}(x,t) P_{\mu}(x,t) P_{\nu}(x,t) \rangle^{r=0}_{g=0}
=\frac{C^{\lambda^*}_{\mu \nu}(t)}{b_{\lambda}(t)} ,  \quad \lambda, \mu, \nu \in \mathcal{P}_{N, \kappa}. 
\end{align}
Now, we are ready to express the genus zero two point functions of $\{ P_{\lambda}(x,t) \}_{\lambda \in \mathcal{P}_{N, \kappa}}$ 
for $r=2$ as a linear combination of genus zero three point  functions for $r=0$.
From \eqref{eq:corCSadj}, we have the following relation 
\begin{align}
\langle  P_{\mu}(x,t) P_{\nu}(x,t)  \rangle^{r=2}_{g=0}=(1-t)^{2N} \langle  P_{\mu}(x,t) P_{\nu}(x,t)  \Delta(x,t)^2 \rangle^{r=0}_{g=0}\,,
\end{align}
where we defined
\begin{align}
\Delta(x,t):=\prod_{a, b =1 \atop a \neq b}^N (1-t x_a x_b^{-1}) \label{saseki}.
\end{align}
$\Delta(x,t)^2$ is also written as
\begin{align}
\Delta(x,t)^2= \left(\prod_{c=1}^{N} x^{2(1-N)}_c \right) \prod_{a \neq b}^N(x_b-t x_a )^2\,.
\label{eq:diff1}
\end{align}
Since   the factor $\prod_{c=1}^{N} x^{2(1-N)}_c$ on the right hand side of \eqref{eq:diff1}
 is always rewritten as a symmetric monomial by using the relation  $\prod_{a=1}^N x^{\kappa}_a \equiv 1$  
 which follows from \eqref{eq:AppB1} with $\lambda=(\kappa, \cdots, \kappa)$, 
we find that \eqref{eq:diff1}  is equal to a symmetric polynomial of $x$ up to the Bethe ansatz equation    
and can be expanded by $\{ P_{\lambda}(x,t) \}_{\lambda \in \mathcal{P}_{N, \kappa}}$ as
\begin{align}
\Delta(x,t)^2 \equiv \sum_{\lambda \in \mathcal{P}_{N, \kappa}} g_{\lambda}(t) P_{\lambda}(x,t)\,.
\label{eq:difff}
\end{align}
Thus the genus zero two point functions for $r=2$  are written by the structure constant $C^{\lambda}_{\mu \nu}(t)$ as
\begin{align}
\eta_{\mu \nu }&:=\langle P_{\mu} P_{\nu}  \rangle^{r=2}_{g=0} 
\equiv (1-t)^{2N}  \sum_{\lambda \in \mathcal{P}_{N, \kappa}} g_{\lambda}(t) \langle P_{\lambda} P_{\mu} P_{\nu}   \rangle^{r=0}_{g=0} \nonumber \\
&= (1-t)^{2N}  \sum_{ \lambda \in \mathcal{P}_{N, \kappa}} \frac{g_{\lambda}(t) C_{\mu \nu}^{\lambda^*}(t)}{b_{\lambda}(t)}   \,.
\label{eq:3pointfunc}
\end{align}
On the right hand side of \eqref{eq:3pointfunc}, the dependence of $R$-charge only comes from $(1-t)^{2N}$ and  $ g_{\lambda}(t)$. 
In Appendix C we discuss properties of the metric and the coupling in the case of general $R$-charge.  In particular 
we will show that we have 2d TQFT for any integral charge $r$. 
In the following subsections, we will evaluate genus $g$ partition functions of $U(2)$ with level $\kappa=2,3,4$,  
$U(3)$ with level $\kappa=2,3$ and $U(4)$ with level $\kappa=2$ from \eqref{eq:deffusion} and \eqref{eq:3pointfunc}.

\subsection{$U(2)$ cases}
 First we rewrite the insertion factor $\Delta(x, t)^2$ in terms of Hall-Littlewood polynomials which holds for general $\kappa \ge 2$. 
When $N=2$, the insertion factor is
\beq
\Delta(x, t)^2 
= e_2(x)^{-2} \left( (1+t)^2 e_2(x) -t (e_1(x))^2 \right)^2,
\eeq
where $e_\ell(x)$ is the elementary symmetric polynomial of degree $\ell$.
Using the relation $e_N(x)^\kappa \equiv1$ which comes from the Bethe ansatz equation,
we may evaluate
\beq
\Delta(x, t)^2 \equiv e_2(x)^{\kappa-2} \left( (1+t)^2 e_2(x) -t (e_1(x))^2 \right)^2.
\eeq
The relation of the elementary symmetric polynomials and the Schur function 
is $e_\ell(x) = s_{(1^\ell)}(x)$. Hence we have
\beq
\Delta(x, t)^2 \equiv s_{(1,1)}^{\kappa-2} \left((1+t)^2 s_{(1,1)}-t (s_{(1)})^2 \right)^2.
\eeq
From now on we do not write $x$ dependence explicitly. 
From the composition rule of two $SU(2)$ representations or two spins,
we see
\beq
s_{(1)}^2 = s_{(2)} + s_{(1,1)}, \qquad s_{(2)}^2 = s_{(4)} + s_{(3,1)} + s_{(2,2)},
\eeq
which gives
\begin{align}
\Delta(x, t)^2 &\equiv s_{(1,1)}^{\kappa-2}
\left( t^2 s_{(4)} - (2t + t^2 + 2t^3) s_{(3,1)} + (1+2t + 4 t^2 + 2 t^3 +t^4) s_{(2,2)} \right) \CR
&= t^2 s_{(\kappa+2, \kappa-2)} - (2t + t^2 + 2t^3) s_{(\kappa+1,\kappa-1)} + (1+2t + 4 t^2 + 2 t^3 +t^4) s_{(\kappa,\kappa)}.
\end{align}
We want to emphasize that this is a universal formula valid for any level $\kappa$.
When $N=2$ the relation of the Schur functions and Hall-Littlewood polynomials is\footnote{This is {\it not} true
for $N>2$, since there appears the partition of length greater than two. We have truncated 
the transition matrix in Macdonald's book \cite{Macdonald:book} by the partitions up to length two.}
\beqa
s_{(\kappa+2, \kappa-2)} &=& P_{(\kappa+2, \kappa-2)} + t P_{(\kappa+1, \kappa-1)} + t^2 P_{(\kappa, \kappa)}, \CR
s_{(\kappa+1, \kappa-1)} &=& P_{(\kappa+1, \kappa-1)} + t P_{(\kappa, \kappa)}, \\
s_{(\kappa, \kappa)} &=& P_{(\kappa, \kappa)}. \nonumber
\eeqa
Hence, we arrive at
\beq
\Delta(x, t)^2 \equiv 
t^2 P_{(\kappa+2, \kappa-2)} - (2t + t^2 + t^3) P_{(\kappa+1,\kappa-1)} + (1+2t + 2 t^2 +  t^3) P_{(\kappa,\kappa)}.
\label{eq:Delta2U2}
\eeq

The next task is to express the right hand side in terms of the Hall-Littlewood polynomials 
in the fundamental domain $\mathcal{P}_{2, \kappa}$.
Here the ideal $\mathcal{I}_{N,\kappa}$ in the ring of the symmetric polynomials 
$\Lambda_N = \R [ x_1, \cdots x_N]^{\mathfrak{S}_N}$
depends on the level $\kappa$ and we have to consider case by case with level $\kappa$. 
As discussed by Korff in \cite{Korff:2013rsa} we can obtain any weight $\omega = \sum_{i=1}^N \omega_i \epsilon_i$ in the $\mathfrak{gl}_N$ 
weight lattice $\Z [\epsilon_1, \cdots \epsilon_N]$ from an appropriate 
element in $\mathcal{P}_{N, \kappa}$ (hence the name \lq\lq fundamental domain\rq\rq) 
by the action of the affine Weyl group $\widetilde{\mathfrak{S}}_{N,\kappa}$ 
with level $\kappa$, which includes the translation of length $\kappa$ in addition to the usual permutations. 
We can obtain the necessary relations among the Hall-Littlewood polynomials involved
in the process of the action of $\widetilde{\mathfrak{S}}_{N,\kappa}$. 

When $\kappa=2$ we have 
\beq
\Delta(x, t)^2 \equiv 
t^2 P_{(4,0)} - (2t + t^2 + t^3) P_{(3,1)} + (1+2t + 2 t^2 +  t^3) P_{(2,2)}.
\eeq
The first two terms are outside $\mathcal{P}_{2, 2}$, while the last term is already in  $\mathcal{P}_{2, 2}$.
We note $\lambda \cdot \sigma_0 = (4,0)$ for $\lambda = (2,2)$ and 
$\lambda \cdot \tau = (3,1)$ for $\lambda = (1,1)$. 
The definition of the actions of $\sigma_0$ and $\tau$ is given in Appendix B. 
Using \eqref{eq:AppB2} and \eqref{eq:AppB5} with the relation of $R_\lambda$ and $P_\lambda$, we see
\beq
P_{(4,0)} \equiv t(1+t) P_{(2,2)} + (t-1) P_{(3,1)}, \qquad
P_{(3,1)} \equiv (1+t) P_{(1,1)}.
\eeq
Substituting them, we finally obtain
\beq
\Delta(x,t)^2 \equiv (1+t)^2 \left( (1+t^2) P_{(2,2)} -2t P_{(1,1)} \right).
\label{eq:gU2k2}
\eeq

For $\kappa=3$ we have 
\beq
\Delta(x, t)^2 \equiv 
t^2 P_{(5,1)} - (2t + t^2 + t^3) P_{(4,2)} + (1+2t + 2 t^2 +  t^3) P_{(3,3)}.
\eeq
As before the first two terms are outside $\mathcal{P}_{2, 3}$, while the last term is already in  $\mathcal{P}_{2, 3}$.
From $\lambda = (2,1)$, we can obtain $\lambda\cdot \tau = (4,2)$ and $\lambda\cdot \sigma_1 \cdot \tau = (5,1)$,
which implies 
\beq
P_{(4,2)} \equiv P_{(2,1)}, \qquad P_{(5,1)} \equiv t P_{(2,1)}.
\eeq
Substituting them, we finally obtain
\beq
\Delta(x,t)^2 \equiv (1+t)(1+t+t^2) P_{(3,3)} -t(2+t) P_{(2,1)}.
\eeq

For $\kappa=4$ we have
\beq
P_{(6, 2)} \equiv (1+t) P_{(2, 2)}, \qquad P_{(5, 3)} \equiv P_{(3, 1)},
\eeq
which lead us to 
\beq
\Delta(x,t)^2 \equiv 
t^2 (1+t) P_{(2, 2)} - (2t + t^2 + t^3) P_{(3,1)} + (1+t)(1+t+t^2) P_{(4,4)}.
\eeq

When $\kappa \geq 5$ we see a phenomenon of \lq\lq stabilization\rq\rq\ in the semi-classical limit $\kappa \to \infty$.  
Namely by acting $\tau$ we observe  
\beq
P_{(\kappa+2, \kappa-2)} \equiv P_{(\kappa-2, 2)}, \qquad P_{(\kappa+1, \kappa-1)} \equiv P_{(\kappa-1, 1)}\,.
\eeq
Thus we obtain a general formula for $\kappa \geq 5$;
\beq
\Delta(x,t)^2 \equiv 
t^2 P_{(\kappa-2, 2)} - (2t + t^2 + t^3) P_{(\kappa-1,1)} + (1+t)(1+t+t^2) P_{(\kappa,\kappa)}.
\eeq

By using the formula of $\Delta (x,t)^2$, we can write down 
genus zero partition functions in the $U(2)_{\kappa}$ models
\begin{eqnarray}
&&Z_{g=0}=C_{\emptyset\emptyset\emptyset}(t)=(1-t)^4\frac{g_{\emptyset}(t)}{b_{\emptyset}(t)}=
\left\{
\begin{array}{ll}
(1-t)(1-t^4) & (\kappa =2)\\
(1-t)(1-t^3) & (\kappa >2)
\end{array}
\right.
\,.
\end{eqnarray}

\subsubsection*{\underline{Level $\kappa=2$}}
We explain how to calculate $C^{\lambda}_{\mu \nu}$ for $\kappa=2$ in detail .
We fix the order of elements of $\mathcal{P}_{2, 2}$  to use matrix notation as follows
\begin{align}
\mathcal{P}_{2, 2}=\{ (2,2), (2,1), (1,1) \}\,.
\label{eq:A22}
\end{align}
When $\kappa=2$ for general $N$ the $*$-involution of an element of $\lambda \in \mathcal{P}_{2, N}$ is same as itself;  $\lambda^* =\lambda$.
Since  $P_{(2,2)}(x,t)\equiv1$ for $x \in \mathrm{Sol}$, we have  
relations 
\begin{align}
&P_{(2,2)} P_{(2,2)}  \equiv P_{(2,2)}, 
\label{eq:CU2k2no1}
\\
&P_{(2,2)} P_{(2,1)} \equiv P_{(2,1)},
\label{eq:CU2k2no2} \\
&P_{(2,2)} P_{(1,1)} \equiv P_{(1,1)}.
\label{eq:CU2k2no3}
\end{align}
When $x=(x_1, x_2)$ is a set of generic variables which does not satisfy the Bethe ansatz equation, 
products of Hall-Littlewood polynomials are expanded as
\begin{align}
&P_{(2,1)} P_{(2,1)}= P_{(4,2)}+(1+t) P_{(3,3)} , \\
&P_{(2,1)} P_{(1,1)}= P_{(3,2)}, \\
&P_{(1,1)} P_{(1,1)} = P_{(2,2)}.
\end{align} 
From an identity \eqref{eq:AppB2} for $(4,2)= (2,2) \cdot \tau$ and $(3,2)= (2,1) \cdot \tau$, 
polynomials 
 $P_{(4,2)}, P_{(3,3)}, P_{(3,2)}$ can be expressed as combinations of  $ \{ P_{\lambda} \}_{\lambda \in \mathcal{P}_{2,2}}$.  Then we have
\begin{align}
&P_{(2,1)} P_{(2,1)} \equiv (1+t) P_{(2,2)}+(1+t) P_{(1,1)}, \\
&P_{(2,1)} P_{(1,1)} \equiv P_{(2,1)}.
\end{align}
The structure constants 
$C^{\lambda}_{\mu \nu} (t)$'s for $N=2, \kappa=2$ in the matrix notation  are given by
\begin{align}
C^{(2,2)}_{\mu \nu}=\left(
\begin{array}{ccc}
 1 & 0 & 0 \\
 0 & 1+t & 0\\
 0 & 0 & 1\\
\end{array}
\right), \,
C^{(2,1)}_{\mu \nu}=\left(
\begin{array}{ccc}
 0 & 1 & 0 \\
 1 & 0 & 1\\
 0 & 1 & 0\\
\end{array}
\right), \,
C^{(1,1)}_{\mu \nu}=\left(
\begin{array}{ccc}
 0 & 0 & 1 \\
 0 & 1+t & 0\\
 1 & 0 & 0\\
\end{array}
\right).
\label{eq:fusionU2k2}
\end{align}
From  \eqref{eq:gU2k2} and \eqref{eq:fusionU2k2}, the metric is given by
\begin{align}
&\eta_{ \mu \nu}=\left(
\begin{array}{ccc}
 (1-t)^2 (1+t) (1+t^2) & 0 & -2  t(1-t)^2 (1+t) \\
 0 & (1-t)^2 (1-t^2)^2 & 0 \\
 -2  t (1-t)^2 (1+t) & 0 & (1-t)^2 (1+t) (1+t^2) \\
\end{array}
\right).
\end{align}
The characteristic polynomial of the handle operator is defined as 
$\det \left( y I -H \cdot C \right) = (y-y_1)^2(y-y_2)$ with $y_1$ and $y_2$
\begin{align}
y_1=\frac{4}{(1-t)^4(1+t)}, \quad y_2=\frac{2}{(1-t)^2(1+t)^3} \,.
\end{align}
So we can write down the genus $g$ partition function in this model
\begin{align}
&Z^{U(2)_{\kappa=2}}_{g}=\left( \frac{2}{(1-t)^2(1+t)^3} \right)^{g-1}+2 \left( \frac{4}{(1-t)^4(1+t)} \right)^{g-1} \,.
\end{align}

\subsubsection*{\underline{Level $\kappa=3$}}
Next we evaluate the model with level $\kappa=3$, namely, 
$U(2)_{\kappa =3}$ Chern-Simons theory with an adjoint chiral multiplet for $r=2$.
In this case, $\mathcal{P}_{2, 3}$ consists of six partitions
\begin{align}
\mathcal{P}_{2, 3}=\{ (3,3), (3,2), (3,1), (2,2), (2,1), (1,1) \}\,.
\end{align}
and the $*$-involution acts on these six elements 
\begin{align}
& (3,3)^*=(3,3), \, (3,2)^*=(3,1), \, (3,1)^*=(3,2),  \\
& (2,2)^*=(1,1), \, (2,1)^*=(2,1), \, (1,1)^*=(2,2).
\end{align}
By using the relations \eqref{eq:AppB2}-\eqref{eq:AppB1} in Appendix B,  
the structure constants are calculated in similar manner as $\kappa=2$ case.
For example,
\begin{align}
P_{(3,1)}P_{(2,1)}=P_{(5,2)}+(1+t)P_{(4,3)} \equiv  (1+t)P_{(2,2)}+P_{(3,1)}\,.
\end{align}
By using relations  $(5,2) =\lambda \cdot \tau $ for $\lambda=(2,2)$ and $(4,3) =\lambda \cdot \tau $ for $\lambda=(3,1)$, we have   relations
\begin{align}
P_{(5,2)} \equiv (1+t) P_{(2,2) }, \quad 
P_{(4,3)} \equiv P_{(3,1) }\,.
\end{align}
Then  structure constants in matrix notation are  given by 
{ \footnotesize
\begin{align}
&C^{(3,3)}_{\mu \nu}=\left(
\begin{array}{cccccc}
 1 & 0 & 0 & 0 & 0 & 0 \\
 0 & 0 & 1+t & 0 & 0 & 0 \\
 0 & 1+t & 0 & 0 & 0 & 0 \\
 0 & 0 & 0 & 0 & 0 & 1 \\
 0 & 0 & 0 & 0 & 1+t & 0 \\
 0 & 0 & 0 & 1 & 0 & 0 \\
\end{array}
\right), \, 
C^{(3,2)}_{\mu \nu}=\left(
\begin{array}{cccccc}
 0 & 1 & 0 & 0 & 0 & 0 \\
 1 & 0 & 0 & 0 & 1 & 0 \\
 0 & 0 & 1 & 1 & 0 & 0 \\
 0 & 0 & 1 & 0 & 0 & 0 \\
 0 & 1 & 0 & 0 & 0 & 1 \\
 0 & 0 & 0 & 0 & 1 & 0 \\
\end{array}
\right), 
\end{align}
\begin{align}
&C^{(3,1)}_{\mu \nu}=\left(
\begin{array}{cccccc}
 0 & 0 & 1 & 0 & 0 & 0 \\
 0 & 1 & 0 & 0 & 0 & 1 \\
 1 & 0 & 0 & 0 & 1 & 0 \\
 0 & 0 & 0 & 0 & 1 & 0 \\
 0 & 0 & 1 & 1 & 0 & 0 \\
 0 & 1 & 0 & 0 & 0 & 0 \\
\end{array}
\right), \,
C^{(2,2)}_{\mu \nu}=\left(
\begin{array}{cccccc}
 0 & 0 & 0 & 1 & 0 & 0 \\
 0 & 1+t & 0 & 0 & 0 & 0 \\
 0 & 0 & 0 & 0 & 1+t & 0 \\
 1 & 0 & 0 & 0 & 0 & 0 \\
 0 & 0 & 1+t & 0 & 0 & 0 \\
 0 & 0 & 0 & 0 & 0 & 1 \\
\end{array}
\right), 
\end{align}
\begin{align}
&C^{(2,1)}_{\mu \nu}=\left(
\begin{array}{cccccc}
 0 & 0 & 0 & 0 & 1 & 0 \\
 0 & 0 & 1 & 1 & 0 & 0 \\
 0 & 1 & 0 & 0 & 0 & 1 \\
 0 & 1 & 0 & 0 & 0 & 0 \\
 1 & 0 & 0 & 0 & 1 & 0 \\
 0 & 0 & 1 & 0 & 0 & 0 \\
\end{array}
\right), \,
C^{(1,1)}_{\mu \nu}=\left(
\begin{array}{cccccc}
 0 & 0 & 0 & 0 & 0 & 1 \\
 0 & 0 & 0 & 0 & t+1 & 0 \\
 0 & 0 & t+1 & 0 & 0 & 0 \\
 0 & 0 & 0 & 1 & 0 & 0 \\
 0 & t+1 & 0 & 0 & 0 & 0 \\
 1 & 0 & 0 & 0 & 0 & 0 \\
\end{array}
\right).
\end{align}
}
Now we introduce the  characteristic polynomial
\begin{align}
\det \left( y I -H \cdot C \right) = (y-y_+)^3 (y-y_-)^3 \,,
\end{align}
where
\begin{align}
y_{\pm}=\frac{3 \left(4t^3 +9t^2 +9 t + 5 \pm (3t^2 + 5t + 1)\sqrt{4t + 5} \right)}{2(1-t)(1-t^2)^3}\,.
\end{align}
The partition function with genus $g$ can be  described by using $y_{\pm}$
\begin{align}
&Z^{U(2)_{\kappa=3}}_{g}= 3 y^{g-1}_+ + 3 y^{g-1}_- \,.
\end{align}

\subsubsection*{\underline{Level $\kappa=4$}}
We fix the order of elements $\mathcal{P}_{2,4}$ as
\begin{align}
\mathcal{P}_{2, 4}=\{ (4,4), (4,3), (4,2), (4,1), (3,3), (3,2), (3,1), (2,2), (2,1), (1,1)  \}\,.
\end{align}
By computing the structure constants and the metric,
we obtain the characteristic polynomial of the handle operator
\begin{align}
\det \left( y I -H \cdot C \right) = (y-y_1)^4 (y-y_2)^4 (y-y_3)^2\,,
\end{align}
where
\begin{align}
y_{1}=\frac{ 8}{(1-t)^2( 1+t)}, \quad y_{2}=\frac{ 8(t + 3)}{(1-t)^4}, \quad y_3=\frac{2(t + 3)}{(1-t)^2( 1+t)^3}\,.
\end{align}
The partiton function with genus $g$ is represented as 
\begin{align}
&Z^{U(2)_{\kappa=4}}_{g}=4 (y_1^{g-1}+y_2^{g-1})+2y_3^{g-1}  \,.
\end{align}
We have computed the genus $g$ partition functions in $SU(2)$ and $U(2)$ models with $\kappa=2,3,4$ and 
 find  these partitions functions actually satisfy the relation \eqref{UNSUNnorm}
\begin{align}
Z^{SU(2)_{\kappa}}_{g}=\left( \frac{2}{\kappa} \right)^g (1-t)^{g-1} Z^{U(2)_{\kappa}}_{g}  \,.
\end{align}
This result means the $U(2)$  partition functions reproduce
the Coulomb branch limit of  the lens space index on $S^1 \times L(\kappa,1)$ for $A_1$ class $\mathcal{S}$ theories \cite{Gukov:2016lki}.

\subsection{$U(3)$ cases}
Let us calculate partition functions for $\kappa =2, 3$.
When $N=3$, the insertion factor  $\Delta(x,t)$ reduces to 
\begin{align}
&P_{(2,2,2)}\Delta(x,t) =(-t^3)P_{(4,2)}+t^2(1 + t)P_{(4,1,1)}+t^2(1 + t)P_{(3,3)}\nonumber \\ 
&\qquad \qquad  -t(1 + t)P_{(3,2,1)}+(1+t)(1+t+t^2)P_{(2,2,2)} \,.
\end{align}
It is amusing that $t^4, t^5$ and $t^6$ terms become implicit by the use of $t$ dependent 
Hall-Littlewood polynomials.
At this stage we may use the ideal relations that depend on the level $\kappa$. When $\kappa=2$,
the relevant relations from \eqref{eq:AppB2}-\eqref{eq:AppB1} are 
\beqa
P_{(2,2,2)} &\equiv& 1 \,,\CR
P_{(4,1,1)} &\equiv& P_{(1,1,2)} \equiv t P_{(1,2,1)} \equiv t^2 P_{(2,1,1)}\,, \CR
P_{(3,2,1)} &\equiv& (1+t) P_{(2,1,1)}\,, \\
P_{(3,3)} &\equiv& t^2 P_{(2,1,1)}\,, \CR
P_{(4,2)} &\equiv& t(1+t)(1+t+t^2) P_{(2,2)} - (1-t^2) P_{(2,1,1)}\,. \nonumber
\eeqa
which allow us to write
\beq
\Delta(t) \equiv (1-t)(1+t)^2(1+t^2)(1+t + t^2) P_{(2,2,2)} - t(1+t)^3(1-t) P_{(2,1,1)}.
\eeq
On the other hand when $\kappa=3$ we can use relations
\beqa
(1+t) P_{(4,1,1)} &\equiv& (1+t) (1+t+t^2) P_{(1,1,1)}\,, \CR
(1+t) P_{(3,3)} &\equiv& (1+t) (1+t+t^2) P_{(3,3,3)}\,, \\
P_{(4,2,0)} &\equiv& t P_{(3,2,1)}\,, \nonumber
\eeqa
which lead to
\beqa
\Delta(t) &\equiv&
 (1+t) (1 + t + t^2) P_{(2,2,2)} -  t (1+t +t^3) P_{(3,2,1)}   \CR
&& + t^2(1+t+t^2)  P_{(3,3,3)} 
+ t^2 (1+2t )(1+t+t^2)  P_{(1,1,1)}.
\eeqa
Next we shall consider the structure constants  and evaluate partition functions for $\kappa=2, 3$.
%
\subsubsection*{\underline{Level $\kappa=2$}}
$\mathcal{P}_{3,2}$ consists of following four partitions:
\begin{align}
\mathcal{P}_{3, 2}=\{ (2,2,2), (2,2,1), (2,1,1), (1,1,1) \}\,.
\end{align}
The structure constants for $\kappa=2$ are shown explicitly 
{ \footnotesize
\begin{align}
C^{(2,2,2)}_{\mu \nu}=\left(
\begin{array}{cccc}
 1 & 0 & 0 & 0 \\
 0 & 1+t+t^2 & 0 & 0 \\
 0 & 0 & 1+t+t^2 & 0 \\
 0 & 0 & 0 & 1 \\
\end{array}
\right)\,, \, 
C^{(2,2,1)}_{\mu \nu}=\left(
\begin{array}{cccc}
 0 & 1 & 0 & 0 \\
 1 & 0 & 1+t & 0 \\
 0 & 1+t & 0 & 1 \\
 0 & 0 & 1 & 0 \\
\end{array}
\right)\,, \\
C^{(2,1,1)}_{\mu \nu}=\left(
\begin{array}{cccc}
 0 & 0 & 1 & 0 \\
 0 & 1+t & 0 & 1 \\
 1 & 0 & 1+t & 0 \\
 0 & 1 & 0 & 0 \\
\end{array}
\right)\,, \, 
C^{(1,1,1)}_{\mu \nu}=
\left(
\begin{array}{cccc}
 0 & 0 & 0 & 1 \\
 0 & 0 & 1+t+t^2 & 0 \\
 0 & 1+t+t^2 & 0 & 0 \\
 1 & 0 & 0 & 0 \\
\end{array}
\right)\,.
\end{align}
}
Then characteristic polynomial is defined as
\begin{align}
\det \left( y I -H \cdot C \right) = (y-y_{+})^2 (y-y_-)^2\,,
\end{align}
\begin{align}
y_{\pm}&=\frac{1}{(1-t)^9( 1+t)^3 (   1+t+t^2
)^5}\Bigl( (5t^2 + 6t + 5) (t^6 + t^5 + 4t^4 + 4t^3 + 4t^2 + t + 1) \nonumber \\
 & \qquad      \pm (1+t) (t^6 + 5t^5 + 6t^4 + 8t^3 + 6t^2 + 5t + 1) \sqrt{5t^2 + 6t + 5} \Bigr)\,.
 \label{U3level2}
\end{align}
We have  the genus $g$ partition function for  the $U(3)_{\kappa=2}$ CS-matter theory \begin{align}
&Z^{U(3)_{\kappa=2}}_{g}=2 (y_{+}^{g-1}+y_{-}^{g-1})  \,.
\label{eq:genusgU3k2}
\end{align}
We will show several examples in lower genera
\begin{align}
&Z^{U(3)_{\kappa=2}}_{g=0}= (1 - t)^2 (1 -t^2)^3 (1 + t + 4 t^2 + 4 t^3 + 4 t^4 + t^5 + t^6) \,,\\
\label{eq:N3k2g0}
&Z^{U(3)_{\kappa=2}}_{g=1}= 4\,, \\
&Z^{U(3)_{\kappa=2}}_{g=2}=\frac{4 \left(5 t^2+6 t+5\right) \left(t^6+t^5+4 t^4+4 t^3+4 t^2+t+1\right)}{(1-t) (1-t^2)^3 \left(1-t^3\right)^5}\,, \\
&Z^{U(3)_{\kappa=2}}_{g=3}=\frac{8 \left(5 t^2+6 t+5\right)}{(1-t)^{2} (1-t^2)^6 \left(1-t^3 \right)^{10}}
  (3 t^{14}+14 t^{13}+60 t^{12}+152 t^{11}+309 t^{10} \nonumber \\
& \quad \quad  +490 t^9+660 t^8+720 t^7+660 t^6+490 t^5+309 t^4+152 t^3+60 t^2+14
   t+3) \,.
\label{eq:N3k2g4}
\end{align}
\subsubsection*{\underline{Level $\kappa=3$}}
$\mathcal{P}_{3,3}$ consists of following ten partitions
\begin{align}
\mathcal{P}_{3, 3}=\{ (3,3,3), (3,3,2), (3,3,1), (3,2,2), (3,2,1), (3,1,1), (2,2,2), (2,2,1), (2,1,1), (1,1,1)  \}\,.
\end{align}
By computing the structure constants and the metric,
we obtain the characteristic polynomial
\begin{align}
\det \left( y I -H \cdot C \right) = (y-y_1)^6 (y-y_2)^3 (y-y_3) \,,
\end{align}
\beeq
&&y_1= \frac{9}{(1-t)^4(1-t^2)^3
   \left(1-t^3\right)},\,
y_2=\frac{9 (t+2)^2}{(1-t)^9 (t+1)^3},\,
y_3=\frac{(t+2)^2}{(1-t)^3 \left(t^2+t+1\right)^5}. \nonumber \\
\eeeq
In this model, the partition function is given by
\beeq
&&Z_{g}=6y_1^{g-1}+3y_2^{g-1}+y_3^{g-1}\,,
\label{eq:genusgU3k3}
\eeeq
and we show several examples in lower genera
\begin{align}
&Z^{U(3)_{\kappa=3}}_{g=0}=(1-t)^3 \left(t^8+2 t^7+6 t^6+6 t^5+3 t^4+3 t^3+3 t^2+2 t+1\right) \,, \\
&Z^{U(3)_{\kappa=3}}_{g=1}= 10\,, \\
&Z^{U(3)_{\kappa=3}}_{g=2}= \frac{1}{(1-t) (1-t^2)^3 \left(1-t^3\right)^5}
(81 t^{12}+244 t^{11}+1054 t^{10}+2746 t^9+6071 t^8 \nonumber \\
& \qquad \qquad +9503 t^7+11909 t^6+11138 t^5+8513 t^4+4808 t^3+2176 t^2+640 t+166)  \,.
\end{align}
By using the relation \eqref{UNSUNnorm} 
 we can obtain $SU(3)_{\kappa=2,3}$ partition functions from the results \eqref{eq:genusgU3k2} and \eqref{eq:genusgU3k3}, 
\begin{align}
Z^{SU(3)_{\kappa}}_g= \left(\frac{3}{\kappa} \right)^g (1-t)^{g-1} Z^{U(3)_{\kappa}}_g\,. 
\end{align}
We made use of the mathematica notebook file attached to the arXiv version of \cite{Gukov:2016lki}
to compute $SU(3)_{\kappa=2,3}$ partition functions from  the Coulomb branch limit of indices 
on $S^1 \times L(\kappa,1)$ associated with $T_3$ theory.
We have found that our results in the $SU(3)_{\kappa=2,3}$ models
agree completely with \cite{Gukov:2016lki}.

\subsection{$U(4) $ case}
$\mathcal{P}_{4,2}$ consists of following five partitions
\begin{align}
\mathcal{P}_{4, 2}=\{ (2,2,2,2 ), (2,2,2,1), (2,2,1,1), (2,1,1,1), (1,1,1,1) \}\,.
\end{align}
By computing the structure constants and the metric,
we can write down the characteristic polynomial of the handle operator;
\begin{align}
\det \left( y I -H \cdot C \right) = (y-y_1)^2  (y-y_2)^2 (y-y_3) \,,
\end{align}
with
\begin{align}
&y_1=\frac{4 (3t^2 + 2t + 3)}{(1-t)^{16}(1+t)^6 (1+t +t^2)^5} \,,\CR
&y_2=\frac{4}{( 1-t)^{10}( 1+t)^4 ( 1+t^2 )^3 ( 1+t +t^2)^5}\,, \label{U4level2} \\ 
&y_3=\frac{3t^2 + 2t + 3}{(1-t)^8( 1+t)^{10} ( 1+t^2)^7}\,. \nonumber
\end{align}
Thus the $U(4)$ partition function is given by using $y_1$, $y_2$, and $y_3$ 
\begin{align}
&Z^{U(4)_{\kappa=2}}_{g}= 2 ( y^{g-1}_1 +y^{g-1}_2)+  y^{g-1}_3 \,.
\end{align}
We can also evaluate the $SU(4)$ counterpart from this result $Z^{U(4)_{\kappa=2}}_{g}$
\begin{align}
&Z^{SU(4)_{\kappa=2}}_{g}= 2^g (1-t)^{g-1}   \left( 2  y^{g-1}_1 +2 y^{g-1}_2+  y^{g-1}_3 \right)\,. 
\end{align}
We expect that  $SU(4)_{\kappa=2}$ partition functions constructed from this $U(4)$ theory reproduce the
Coulomb branch limit of indices on $S^1 \times L(2,1)$ associated with $T_4$ theory \cite{Gukov:2016lki}.


\section{Discussions on intriguing aspects of the algebra}

\subsection{Recurrence formula in genus}

Recall the genus $g$ partition functions can be constructed by using 
the handle operator 
$Z_g=
{\rm Tr}\left\{(H\cdot C)^{g-1}\right\}
=\sum_{\mu \in {\cal P}_{N,\kappa}}
\left\{(H\cdot C)^{g-1}\right\}_{\mu}{}^{\mu}$
in \eqref{zgenus}.
One can evaluate the partition function $Z_g$
by computing eigenvalues $y_i$ of the matrix $(H\cdot C)$ as \eqref{zg-eigenvalue}.
These eigenvalues are calculated as roots of a characteristec equation in (\ref{charact-poly})
\begin{eqnarray}
&&\det\left[yI-(H\cdot C)\right]
=\prod_{i}(y-y_{i})^{m_{i}}=0
\quad (m_{i}\in {\mathbb Z}_{>0})\,.
\end{eqnarray}
We can also derive the recurrence equation 
of the partition function $Z_g$ as follows;
First we consider the matrix $M=(H\cdot C)$
and look for the minimal polynomial $f(M):=\prod_{i} (M-y_{i}I)^{\tilde{m}_{i}}~(
\tilde{m}_{i}\in {\mathbb Z}_{>0})$ which can be evaluated by using 
roots of the characteristic equation $\det [yI-M]=0$.
In our models 
($U(2)_{\kappa =2,3,4}$,
$U(3)_{\kappa =2,3}$,
$U(4)_{\kappa =2}$), it turns out 
the minimal polynomials are factorized to first order
polynomials, namely $\tilde{m}_i=1$,
which means the matrix $M=(H\cdot C)$ is diagonalizable,
or there are no Jordan blocks of size greater than one. 
In such cases the Frobenius algebra is called semi-simple. 
\begin{eqnarray}
&&
f(M)=\prod_{i}(M-y_{i}I)=
M^K+b_1M^{K-1}+b_2M^{K-2}+ \cdots +b_{K-1}M+b_KI\,,\nonumber\\
&&K=
\left\{
\begin{array}{cl}
2 & {\rm for}\,\,U(2)_{\kappa =2,3},\,U(3)_{\kappa =2},\\
3 & {\rm for}\,\,U(2)_{\kappa =4},\,U(3)_{\kappa =3},\,U(4)_{\kappa =2}\,.
\end{array}
\right.
\end{eqnarray}
By applying this polynomial equation $f(M=(H\cdot C))=O$ to 
the partition function in our models, we can obtain the recurrence formula
\begin{eqnarray}
&&
Z_{g+K}+b_1Z_{g+K-1}+b_2Z_{g+K-2}+ \cdots +b_{K-1}Z_{g+1}+b_KZ_{g}=0\quad (g=0,1,2,\cdots )\,.\nonumber
\end{eqnarray}

Now we make a comment here:
the partiton function can be evaluated by 
solving the characteristic equation (\ref{charact-poly}) associated with the 
matrix $(H\cdot C)$. 
In applying this procedure to the partiton functions, 
the main difficulty to be encountered 
lies in the evaluation of solutions of the algebraic equation. 
In such situation that we cannot derive the explicit roots 
of the algebraic equations, the recursion equation should be very useful.
So we will explain another derivation of the recursion equation. 

We introduce a generating funtion 
$\mathcal{Z}_{g\geq 2}(s):=\sum_{g\geq 2} s^{g-1}Z_{g}$ of the genus $g$ partiton functions $Z_g$ ($g\geq 2$) 
and rewrite this function
\begin{align}
\mathcal{Z}_{g\geq 2}(s)&={\rm Tr}
\left[s\,(H\cdot C)\cdot \{I-s(H\cdot C)\}^{-1}\right]\nonumber\\
&=-s\frac{d}{ds}{\rm Tr}\log \left[I-s(H\cdot C)\right]
=-\frac{\displaystyle s\frac{d}{ds}F(s)}{F(s)}
\label{zenkasiki1}.
\end{align}
with
\begin{eqnarray}
&&
F(s):=\det \left[I-s(H\cdot C)\right]=
1+\sum_{m=1}^l A_m s^m\,,\\
&&l=\frac{1}{N!}\kappa (\kappa +1)\cdots (\kappa +N-1)\quad {\rm for} \,\, U(N)_{\kappa}\,,
\end{eqnarray}
where the set of coefficients $A_m$'s in $F(s)$
can be expressed by using minors of the matrix $(H\cdot C)$ 
\begin{eqnarray}
&&(H\cdot C)
:=
\left[
\begin{array}{cccc}
a_{1,1} & a_{1,2} & \cdots & a_{1,l} \\
a_{2,1} & a_{2,2} & \cdots & a_{2, l} \\
 \vdots & \vdots & & \vdots  \\
a_{l,1} & a_{l,2} & \cdots & a_{l, l} \\
\end{array}
\right]\,,\\
&&A_m=
(-1)^m \sum_{j_1<j_2<\cdots <j_m}\det
\left[
\begin{array}{cccc}
a_{j_1j_1} & a_{j_1j_2} & \cdots & a_{j_1j_m} \\
a_{j_2j_1} & a_{j_2j_2} & \cdots & a_{j_2j_m} \\
 \vdots & \vdots & & \vdots  \\
a_{j_mj_1} & a_{j_mj_2} & \cdots & a_{j_mj_m} \\
\end{array}
\right]\,.
\end{eqnarray}
Then we can obtain the partition functions recursively by using 
(\ref{zenkasiki1});
\begin{eqnarray}
&&Z_{2}=-A_1\,,\nonumber \\
&&Z_{m+1}=-mA_m-\sum_{k=1}^{m-1}A_kZ_{m+1-k}\quad (m=2,3,\cdots ,l)\,,\\
&&Z_{m+1}=-\sum_{k=1}^{l}A_kZ_{m+1-k}\quad (m=l+1,l+2,l+3 ,\cdots )\,. \nonumber 
\end{eqnarray}
We show several examples in lower genera,
\begin{eqnarray}
&&Z_{g=2}=-A_1\,,\,\,
Z_{g=3}=-2A_2+A_1^2\,,\nonumber \\
&&Z_{g=4}=-3A_3+3A_1A_2-A_1^3\,,\\
&&Z_{g=5}=-4A_4+2A_2^2+4A_3A_1-4A_1^2A_2+A_1^4\,.\nonumber
\end{eqnarray}


\subsection{Fate of level-rank duality}

One of the most significant properties of the Verlinde algebra 
is the level-rank duality \cite{Nakanishi:1990hj, Naculich:1990pa, Mlawer:1990uv,
Naculich:2007nc, Hsin:2016blu}. Unfortunately it seems that 
this duality cannot survive after the $t$-deformation of the algebra. 
We have obtained the explicit forms of genus $g$ partition function $Z_g$
by solving the characteristic equation for the handle operator
for several models. 
We observe the level-rank duality at $t=0$ is 
realized as the agreement of the genus $g$ partition functions
between $SU(N)_\kappa$ and $U(\kappa)_N$ as follows;
\begin{align}
&
Z^{SU(2)_3}_g = Z^{U(3)_2}_{g}=2\cdot 
\left\{(5+\sqrt{5})^{g-1}
+(5-\sqrt{5})^{g-1}\right\}, \, \,  
Z^{U(2)_3}_{g}=\left(\frac{3}{2}\right)^{g}
\cdot Z^{U(3)_2}_{g}, \\
&
Z^{SU(2)_4}_g =Z^{U(4)_2}_{g}=
3^{g-1}+2\cdot 4^{g-1}
+2\cdot 12^{g-1},\,\,  
Z^{U(2)_4}_{g}=2^{g}
\cdot Z^{U(4)_2}_{g}.
\end{align}
Note that if we compare $U(N)_\kappa$ and $U(\kappa)_N$,
we need the additional correction factor $(\kappa/N)^g$,
since their Witten indices (the dimensions of the Hilbert space on genus one curve) 
are different. If we compare \eqref{SU2level3} and \eqref{SU2level4} 
with \eqref{U3level2} and \eqref{U4level2} respectively, 
we see the $t$-dependence of the roots of the characteristic 
polynomial of the handle operator is completely different
and there seems to be no simple relations. 

 
\subsection{Selection Rules in $U(N)_{\kappa}$ Theory }
 
There is the freedom of the $R$-charge $r$ of the adjoint matter in our model ($U(N)_{\kappa}$ theory).  
As can be seen from the general formula \eqref{eq:corCSadj}, the dependence of the $R$-charge $r$
only comes from the difference product $\Delta(x,t)$ defined by \eqref{saseki} and hence the coefficients $\{g_{\lambda}(t)\}$ in the expansion 
\eqref{eq:difff} play an important role in analyzing the $r$-dependence of the model. 
In order to investigate properties of the coefficients $\{g_{\lambda}(t)\}$, 
we will work out selection rules of these coefficients.

First we decompose the set of partitions ${\cal P}_{N,\kappa}$ into 
$\kappa$ subsets ${\cal P}^{(n)}_{N,\kappa}$ $(n=0,1,2,\cdots ,\kappa -1)$ 
according to the number of boxes $|\lambda|:=\displaystyle{\sum_{i=1}^N} \lambda^i$ 
of a partition $\lambda$;
\begin{eqnarray}
&&{\cal P}_{N,\kappa}
=\{\lambda =(\lambda^1,\lambda^2,\cdots ,\lambda^N)\,|\,
\kappa\geq \lambda^1\geq \lambda^2\geq \cdots \geq\lambda^N\geq 1\}
=\bigcup_{n=0}^{\kappa -1}{\cal P}^{(n)}_{N,\kappa}\,,\nonumber\\
&&{\cal P}^{(n)}_{N,\kappa}:=\{\lambda\in {\cal P}_{N,\kappa}\,|\,\,
|\lambda|\equiv n \mod\kappa \}\quad (n=0,1,2,\cdots ,\kappa -1)\,.
\end{eqnarray}
The Hall-Littlewood polynomials are symmetric homogeneous polynomials of Bethe roots $x_a$
and their degrees are given by $|\lambda|$. 
 We shall take a phase transformation on  Bethe roots  $x_a\rightarrow e^{i\theta}x_a$, $\theta\in {\mathbb R}$ ($a=1,2,\dots ,N$).
Then the degree $|\lambda|$ of the polynomial can be read from the 
phase induced from this transformation: $P_{\lambda}(e^{i\theta}x,t)=e^{i\theta |\lambda|}P_{\lambda}(x,t)$.
For example, $P_{(1,1,\cdots ,1)}=x_1x_2\cdots x_N$ is a homogeneous polynomial with degree $N$.

Next we introduce a new polynomial $J(x,t)$ defined by 
\begin{eqnarray}
J(x,t):=\{P_{(1,1,\cdots ,1)}\}^{\kappa M}\prod_{a,b =1}^N(1-tx_ax^{-1}_b)\qquad (\kappa M-(N-1)>0\,,\,\,M\in {\mathbb Z})
\end{eqnarray}
in order to study properties of $\Delta (x,t)$.
 The difference product  $\Delta (x,t)$ is invariant under the transformation $x_a\rightarrow e^{i\theta}x_a$, 
but $J(x,t)$ is a polynomial with degree $\kappa MN$.
Let us expand $\{J(x,t)\}^r$ ($r=0,1,2,3\cdots $) in terms of the Hall-Littlewood polynomials $P_{\mu}(x,t)$;
\begin{eqnarray}
&&\{J(x,t)\}^r=\sum_{|\mu|=\kappa MNr}a^{(r)}_{\mu}P_{\mu}(x,t)\,.
\end{eqnarray}
Because $\{J(x,t) \}^r$ is a homogeneous polynomial with degree $(\kappa MNr)$,  
the partitions  that appear on the right hand side 
should satisfy $|\mu|=\kappa MNr$, namely, ${\mu}\not\in {\cal P}_{N,\kappa}$ in general.
But one can rewrite the expansion by using a set of relations summarized in Appendix B  
and restrict the sum of partitions  to $\lambda\in {\cal P}_{N,\kappa}$, 
$\{J(x,t)\}^r\equiv (1-t)^{rN} \sum_{\lambda\in {\cal P}_{N,\kappa}}g_{\lambda}^{(r)}P_{\lambda}(x,t)$.
In this reduction, one has to use a set of operations $\tau$, $\sigma_i$, $\sigma_0$ acting on 
$\lambda$ \cite{Korff:2013rsa} (See also Appendix B).
In addition, there is an identity $P_{(\kappa ,\kappa ,\cdots ,\kappa )}=1$ due to the set of Bethe equations \eqref{eq:betheqbosn}.
The number of boxes $|\lambda|$ of partitions is important information in our discussion of selection rules.
While the number of boxes $|\lambda|$ may change under these operations,
we can see these changes $\delta |\lambda|$ are multiples of $\kappa$ 
($\delta |\lambda|=\pm \kappa$ for $\tau$-operation, $\delta |\lambda|=\pm \kappa N$ when one uses the identity 
$P_{(\kappa ,\kappa ,\cdots ,\kappa )}=1$, but $\delta |\lambda|=0$ for $\sigma_i$,$\sigma_0$-operations).
Hence it is natural to define the number of boxes $|\lambda|$ modulo $\kappa$.
Since the degree of $\{J(x,t)\}^r$ is a multiple of $\kappa$,
in the decomposition of $\{J(x,t)\}^r$ there appear only
polynomials $P_{\lambda}(x,t)$ with 
$|\lambda|\equiv 0\mod\kappa$, namely $\lambda\in {\cal P}^{(0)}_{N,\kappa}$;
\begin{eqnarray}
\{J(x,t)\}^r\equiv (1-t)^{r N}\sum_{\lambda \in {\cal P}^{(0)}_{N,\kappa}}g^{(r)}_{\lambda}P_{\lambda}(x,t)
\,.
\end{eqnarray}
This gives a selection rule of the coefficients $\{ g^{(r)}_{\lambda} \}$ of the theory with the $R$-charge $r$. 
We obtain an important result that 
nonvanishing expansion coefficients $g_{\lambda}^{(r)}\neq 0$ should 
appear only in $\lambda\in {\cal P}^{(0)}_{N,\kappa}$;
\begin{eqnarray}
&&
{\cal B}^{(r)}_{N,\kappa}:=\{\lambda\in {\cal P}_{N,\kappa} \,;\,g^{(r)}_{\lambda}\neq 0\}
\rightarrow {\cal B}^{(r)}_{N,\kappa}\subset{\cal P}^{(0)}_{N,\kappa}\,. \label{selection}
\end{eqnarray}
It is possible that the size of the set ${\cal B}^{(r)}_{N,\kappa}$ depends on the $R$-charge $r$
and it is desirable to find a criterion for the equality in \eqref{selection}.

Next let us study why the $U(1)$ phase symmetry $e^{i\theta}x_a$ reduces to the discrete one $\Z_{\kappa}$.
We have done the reduction by using the set of Bethe equations
(\ref{eq:betheqbosn}).
When one performs  the $U(1)$ transformation, 
 the phases $e^{i\kappa\theta}$ appear on the left hand sides
in these equations and the set of equations is not invariant in general.
However, if the parameter $\theta$ satisfies the condition $\theta =\frac{2\pi}{\kappa}m$ ($m\in \Z$), 
these Bethe equations are invariant. It is the reason 
why the $U(1)$ symmerty reduces to $\Z_{\kappa}$.

As an application of this $\Z_{\kappa}$ charge, 
we can obtain selection rules for the  couplings in  $U(N)_{\kappa}$ theory.
The fusion couplings $C^{\lambda}_{\mu \nu}$ are defined as the structure constants of Hall-Littlewood polynomials
\eqref{eq:deffusion}.
In the case of  $r=0$, 
the three point functions $C^{(r=0)}_{\mu\nu\lambda}$ in \eqref{sym3} are given by using metric $\eta^{(r=0)}_{\mu\nu}$
in \eqref{eta0}.
By using conservation of  $\Z_{\kappa}$ charge associated with partitions, 
we can write down the following selection rules
\begin{eqnarray}
&&(|\mu|+|\nu|\not\equiv |\lambda|\mod\kappa)\rightarrow C_{\mu\nu}^{\lambda}= 0 \,,\,\,\nonumber\\
&& (|\mu|+|\nu|\not\equiv 0\mod\kappa)\rightarrow \eta^{(r=0)}_{\mu\nu}= 0\,,\nonumber\\
&& (|\mu|+|\nu|+ |\lambda|\not\equiv 0\mod\kappa)\rightarrow C^{(r=0)}_{\mu\nu\lambda}= 0\,,\quad\nonumber\\
&&|\lambda^{\ast}|\equiv -|\lambda|\mod \kappa. \nonumber
\end{eqnarray}

Next let us investigate the case of general $R$-charge $r$.
In this case, three point functions  $C^{(r)}_{\mu\nu\rho}$'s in (\ref{r-sym3}) and metric $\eta^{(r)}_{\mu\nu}$ 
in (\ref{eta-r}) are defined 
by using $g_{\lambda}^{(r)}(t)$ in (\ref{expansion}) and  $C^{\rho}_{\mu\nu}$.
By using conservation of $\Z_{\kappa}$ charge, we can write the following  selection rules
\begin{eqnarray}
&&(\lambda \not\equiv 0\mod\kappa)\rightarrow g^{(r)}_{\lambda}= 0 \,,\nonumber\\
&&(|\mu|+|\nu|+ |\lambda|\not\equiv 0\mod\kappa)\rightarrow C^{(r)}_{\mu\nu\lambda}= 0 \,,\nonumber\\
&&(|\mu|+|\nu|\not\equiv 0\mod\kappa)\rightarrow \eta^{(r)}_{\mu\nu}= 0\,,\nonumber\\
&& (|\mu|+|\nu|\not\equiv 0\mod\kappa)\rightarrow \eta_{(r)}^{\mu\nu}= 0\,.\nonumber
\end{eqnarray}

We arrange the couplings 
 $C^{\nu}_{\lambda\mu}$ into matrices $C_{\lambda}$ whose  components are given as 
$(C_{\lambda})_{\mu}{}^{\nu}=C_{\lambda\mu}^{\nu}$.
In the case $|\lambda|\not\equiv 0\mod\kappa$, trace part $H_{\lambda}$ of the matrix $C_{\lambda}$ 
vanishes, 
$H_{\lambda}:={\rm Tr}~C_{\lambda}=\sum_{\mu\in {\cal P}_{N,\kappa}}C_{\lambda\mu}^{\mu}=0$ 
and $H^{\lambda}=\sum_{\rho}\eta_{(r)}^{\lambda\rho}~{\rm Tr}~C_{\rho}=0$ 
by using $\eta^{\lambda\rho}_{(r)}=0$ for 
$|\lambda|+|\rho|\not\equiv 0\mod\kappa$.

Now we shall investigate properties of partition functions 
by using conservation of $\Z_{\kappa}$ charge. 
 The genus $g$ partition functions $Z_g$ are constructed by using 
 the handle operator
$(H\cdot C)_{\mu}{}^{\nu}=\sum_{\lambda\in {\cal P}_{N,\kappa}}H^{\lambda}(t)C_{\lambda\mu}^{\nu}$ 
and 
they are combined into a generating function $\mathcal{Z}_{g\geq 2}(s)$;
\begin{eqnarray}
&&Z_g=
\sum_{\mu ,\nu\in {\cal P}^{(0)}_{N,\kappa}}H^{\mu}(t)
\left\{(H\cdot C)^{g-2}\right\}_{\mu}{}^{\nu}H_{\nu}(t)
\qquad (g=2,3,4,\cdots)\,,\nonumber\\
&&\mathcal{Z}_{g\geq 2}(s)=\sum_{g\geq 2}s^{g-1}Z_{g}
=
\sum_{\mu ,\nu\in {\cal P}^{(0)}_{N,\kappa}}s\, H^{\mu}(t)
\left[\{I-s(H\cdot C)\}^{-1}\right]_{\mu}{}^{\nu}H_{\nu}(t)\label{pole1}\,.
\end{eqnarray}
When we use the fact $H^{\lambda}(t)=0$ for $\lambda\not\in {\cal P}^{(0)}_{N,\kappa}$, 
we find that 
only $(H\cdot C)_{\mu}{}^{\nu}$ with $\mu ,\nu\in {\cal P}^{(0)}_{N,\kappa}$ can contribute to  
the partition functions $Z_g$.
So we introduce a minor $\widehat{(H\cdot C)}_{\mu}{}^{\nu}:=(H\cdot C)_{\mu}{}^{\nu}$ ($\mu ,\nu\in {\cal P}^{(0)}_{N,\kappa}$)
and denote their eigenvalues as $\{\hat{y}_{i}\}$. (We also write the multiplicity of each eigenvalue $\hat{y}_{i}$ as 
$\hat{m}_{i}\in \Z_{>0}$).
The generating function $\mathcal{Z}_{g\geq 2}(s)$ in \eqref{pole1} is a function of the variable $s$  
and the pole structure  is determined by 
 $\{\hat{y}_{i}\}$. On the other hand, we have another expression of the partition function
$Z_g
=
{\rm Tr}\left\{(H\cdot C)^{g-1}\right\}
=\sum_{\mu \in {\cal P}_{N,\kappa}}
\left\{(H\cdot C)^{g-1}\right\}_{\mu}{}^{\mu}$.
These partition functions  are expressed by using eigenvalues of the handle operator 
and are combined into the generating function 
\begin{eqnarray}
\mathcal{Z}_{g\geq 2}(s)
=\sum_{g\geq 2}s^{g-1}Z_{g}
=\sum_{i}\frac{m_{i}sy_{i}}{1-s y_{i}}\,.\label{pole2}
\end{eqnarray}
Because our models are 
topological field theories, 
two results from \eqref{pole1} and \eqref{pole2} should agree. 
 By comparing the structure of poles in  these equations
we find  the set of eigenvalues  $\{\hat{y}_{i}\}$ of $\widehat{(H\cdot C)}$
and  $\{y_{i}\}$ of $(H\cdot C)$ should  match. 
When one uses this fact, one can obtain the set of eigenvalues $\{y_{i}\}$ of 
$(H\cdot C)$ by analysing the minor $\widehat{(H\cdot C)}$ and its eigenvalues 
$\{\hat{y}_{i}\}$.
 But the multiplicities of eigenvalues are not equal in general. 
It means that the essential properties of partition functions are determined 
by structure of the sector with vanishing $\Z_{\kappa}$ charge.


\section*{Acknowledgments}
The work of H.K. is supported  by Grants-in-Aid for Scientific Research
 (\# 15H05738, \# 18K03274)  from JSPS. 
The work of Y.Y. is supported by JSPS Grant-in-Aid (S), No.16H06335  
and also by World Premier International Research Center Initiative (WPI), MEXT Japan.


\appendix

\section{2d TQFT and Frobenius algebra}

In 2d TQFT there exists a coboundary operator $Q$ that satisfies $Q^2 =0$ and we consider the cohomology 
of $Q$. If 2d TQFT is obtained from a supersymmetric theory in 2 dimensions, $Q$ is one of the generators 
of supersymmetry, which becomes a scalar charge after twisting. In the $U(N)$ or $SU(N)$ theories discussed in this paper,
the equivalence classes of $Q$-closed operators form a finite dimensional commutative Frobenius algebra 
$\mathcal{A}$ which is realized as a quotient of the Weyl invariant Laurent polynomial ring
\bel
\mathcal{A}=\mathcal{R} [x_1,\cdots, x_N, x^{-1}_1,\cdots, x^{-1}_N]^{W(G)}/\mathcal{I},
\label{eq:Frobalg} 
\ee
where $\mathcal{I}$ is the ideal generated by the saddle point equation and 
the coefficient ring $\mathcal{R}$ is generated by flavor Wilson loops.
In $\mathcal{N}=2$ Chern-Simons theories
with a single adjoint matter the flavor symmetry is $U(1)$ and $\mathcal{R} = \mathbb{Z}[[t]]$ can be
identified with the ring of formal power series in the $U(1)$ equivariant parameter $t$. 
Precisely speaking, all the rational functions of $t$ in this paper should be expanded 
as a formal power series around $t=0$. 
The ideal $\mathcal{I}$ is generated by the saddle point equation from which the factor $x^{\alpha} -1$ is removed.

If we choose a basis $\{ \mathcal{O}_{\mu} \}_{\mu \in L}$ of  the algebra $\mathcal{A}$ of dimension $|L|$, 
 the product of $Q$-closed operators  is expanded in  this basis
\bel
\mathcal{O}_{\mu} \mathcal{O}_{\nu} =\sum_{\lambda \in L} C^{\lambda}_{\mu \nu} \mathcal{O}_{\lambda}\,.
\label{eq:Fusion}
\ee
Here we emphasize that the structure constants $C^{\lambda}_{\mu \nu} \in \mathcal{R}$ do not depend on 
$R$-charges, flavor-flavor, gauge-$R$-symmetry CS levels,
because the saddle point equations, and hence the ideal $\mathcal{I}$, are independent of these parameters. 
The Frobenius algebra $\mathcal{A}$  is equipped with a non degenerate bilinear form called metric 
$\eta: \mathcal{A} \otimes \mathcal{A} \to \mathcal{R}$, which is defined by 
the genus zero two point function\footnote{ The precise form of $\mathcal{R}$ depends on theories.}.
In terms of the basis $\{ \mathcal{O}_{\mu} \}$, the metric is written as   
\begin{align}
\eta_{\mu \nu}=\eta (\mathcal{O}_{\mu}, \mathcal{O}_{\nu}) :=\langle  \mathcal{O}_{\mu} \mathcal{O}_{\nu}  \rangle_{g=0} \,.
\label{eq:metric}
\end{align}
We also define ${\eta}^{ \mu \nu}$ as the inverse matrix of $ {\eta}_{\mu \nu}$ which corresponds to a sphere with two right oriented holes. 
Then the bilinear form $\eta^{\mu \nu}$ and the structure constants $C^{\lambda}_{\mu \nu}$ satisfy the following relation;
\begin{align}
 C^{\lambda}_{ \, \, \mu \nu } = \sum_{\rho \in L}   C_{ \mu \nu \rho} {\eta}^{\rho \lambda },
\end{align}
where $ C_{\lambda \mu \nu}$ is the genus zero three point function;
\begin{align}
 C_{\lambda \mu \nu} :=  \langle  \mathcal{O}_{\lambda} \mathcal{O}_{\mu} \mathcal{O}_{\nu}   \rangle_{g=0}.
\end{align}

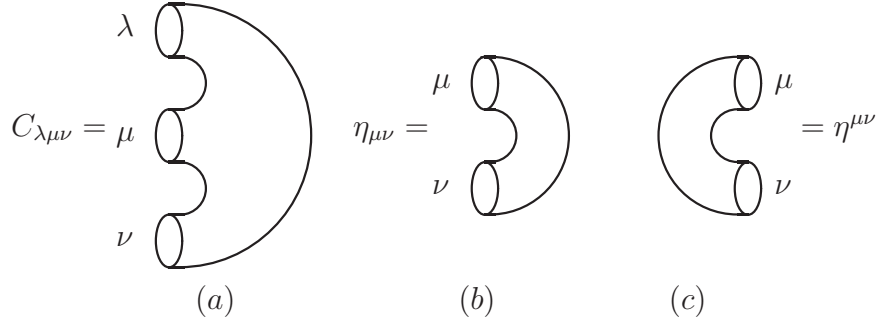
\begin{figure}[t]
\begin{center}
\begin{picture}(70,90)(-70,0)
\unitlength 0.7mm
\thicklines
\put(-65,50){\ellipse{5}{10}}
\put(-65,30){\ellipse{5}{10}}
\put(-65,10){\ellipse{5}{10}}
\put(-65,55){\line(1,0){3}}
\put(-65,45){\line(1,0){3}}
\put(-65,35){\line(1,0){3}}
\put(-65,25){\line(1,0){3}}
\put(-65,15){\line(1,0){3}}
\put(-65,5){\line(1,0){3}}
\put(-63,20){\arc{10}{4.6}{7.8}}
\put(-63,40){\arc{10}{4.6}{7.8}}
\put(-63,30){\arc{50}{4.7}{7.85}}
\put(-60,-3){$(a)$}
\put(-95,30){$C_{\lambda \mu \nu}=$}
\put(-75,49){$\lambda$} 
\put(-75,29){$\mu$} 
\put(-75,9){$\nu$} 

\put(-5,40){\ellipse{5}{10}}
\put(-5,20){\ellipse{5}{10}}
\put(-5,45){\line(1,0){2}}
\put(-5,35){\line(1,0){2}}
\put(-5,25){\line(1,0){2}}
\put(-5,15){\line(1,0){2}}
\put(-4,30){\arc{10}{4.6}{7.8}}
\put(-4,30){\arc{30}{4.6}{7.8}}
\put(-10,-3){$(b)$}
\put(-30,30){$\eta_{\mu \nu}=$}
\put(-15,39){$\mu$} 
\put(-15,19){$\nu$} 

\put(45,40){\ellipse{5}{10}}
\put(45,20){\ellipse{5}{10}}
\put(45,45){\line(-1,0){2}}
\put(45,35){\line(-1,0){2}}
\put(45,25){\line(-1,0){2}}
\put(45,15){\line(-1,0){2}}
\put(43,30){\arc{10}{1.5}{4.8}}
\put(43,30){\arc{30}{1.5}{4.8}}
\put(30,-3){$(c)$}
\put(55,30){$=\eta^{\mu \nu}$}
\put(50,39){$\mu$} 
\put(50,19){$\nu$}

\end{picture}
\end{center}
\caption{$(a)$: $C_{\lambda \mu  \nu }$ corresponds to a sphere with three  left oriented holes. 
$(b)$: $\eta_{\mu \nu}$ corresponds to a sphere with  two  left oriented holes. $(c)$: $\eta^{\mu \nu}$ corresponds to 
a sphere with  two  right oriented holes.}
\label{fig:TFTdata}
\end{figure}

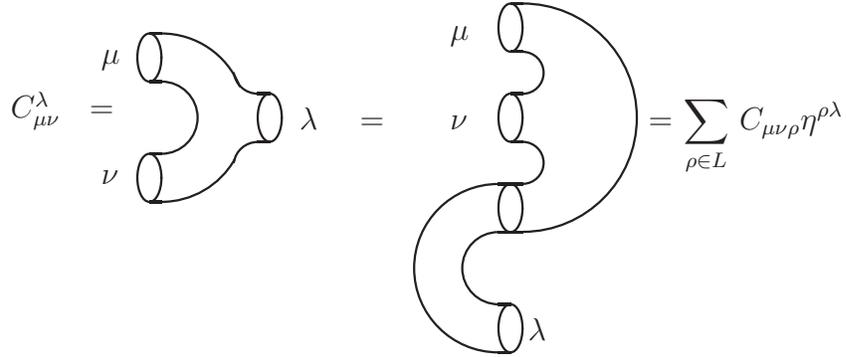
\begin{figure}[t]
\begin{center}
\begin{picture}(90,120)(-30,20)
\unitlength 0.8mm
\thicklines

\put(-35,55){\ellipse{4}{8}}
\put(-35,35){\ellipse{4}{8}}
\put(-35,59){\line(1,0){2}}
\put(-35,51){\line(1,0){2}}
\put(-35,39){\line(1,0){2}}
\put(-35,31){\line(1,0){2}}
\put(-33,45){\arc{12}{4.6}{7.9}}
\put(-33,45){\arc{28}{4.6}{5.9}}
\put(-33,45){\arc{28}{6.8}{7.9}}
\put(-15,49){\line(-1,0){2}}
\put(-15,41){\line(-1,0){2}}
\put(-17,53){\arc{8}{1.5}{3}}
\put(-17,37){\arc{8}{3.3}{4.7}}
\put(-15,45){\ellipse{4}{8}}
\put(25,45){\ellipse{4}{8}}
\put(25,60){\ellipse{4}{8}}
\put(25,30){\ellipse{4}{8}}
\put(25,10){\ellipse{4}{8}}
\put(25,34){\line(-1,0){2}}
\put(25,26){\line(-1,0){2}}
\put(25,14){\line(-1,0){2}}
\put(25,6){\line(-1,0){2}}
\put(23,20){\arc{12}{1.5}{4.8}}
\put(23,20){\arc{28}{1.5}{4.8}}
\put(27,37.5){\arc{7}{4.6}{7.9}}
\put(27,52.5){\arc{7}{4.6}{7.9}}
\put(27,45){\arc{38}{4.7}{7.85}}
\put(25,64){\line(1,0){2}}
\put(25,56){\line(1,0){2}}
\put(25,49){\line(1,0){2}}
\put(25,41){\line(1,0){2}}
\put(25,34){\line(1,0){2}}
\put(25,26){\line(1,0){2}}

\put(-58,45){$C_{\mu \nu}^{\lambda}~~=$}
\put(-43,54){$\mu$}
\put(-43,34){$\nu$}
\put(-10,43){$\lambda$}
\put(0,43){$=$}
\put(15,58){$\mu$}
\put(15,43){$\nu$}
\put(28,8){$\lambda$}

\put(48,43){=~$\displaystyle{\sum_{\rho \in L}}~C_{\mu \nu \rho}\eta^{\rho \lambda}$}
\end{picture}
\end{center}
\caption{The contraction of indices corresponds to gluing holes on Riemann surfaces.}
\label{fig:fusion}
\end{figure}

By definition, $ C_{\lambda \mu \nu}$ is totally symmetric under the permutations of $\lambda, \mu, \nu$ and 
is associated to a sphere with three left oriented holes as figure \ref{fig:TFTdata}.
Note that   $ \eta_{\mu \nu}=C_{\emptyset \mu \nu}$ with $\mathcal{O}_{\emptyset}:=1$.
The specialization  $\mathcal{O}_{\lambda} \to 1$ in the correlation function corresponds to closing a hole with the left orientation 
and the correlator reduces  to the sphere partition function 
with two left oriented holes as figure \ref{fig:TFTdata}.
The contraction of upper and lower indices corresponds to gluing a left hole and a right hole.  
For example,  see figure \ref{fig:fusion}. 
The three point functions have to satisfy  the  associativity condition
\bel
\sum_{\lambda  \in L} C^{\lambda}_{\, \mu \nu } C^{\sigma}_{\, \lambda \rho }=\sum_{\lambda \in L}  C^{\lambda }_{\, \nu \rho } C^{\sigma}_{\, \mu \lambda }\,.
\label{eq:associative}
\ee
The associativity corresponds to  figure \ref{fig:associativity}.
The associativity condition is equivalent  to the existence of $S_{\mu \nu}$ such that
\begin{align}
C^{\lambda}_{\, \mu \nu } =\sum_{\sigma \in L} \frac{S_{\mu \sigma} S_{\nu \sigma } S^{-1}_{\sigma \lambda} }
{S_{\emptyset \sigma}}\,.
\label{Verlinde}
\end{align}
%
%
\begin{figure}[t]
\begin{center}
\begin{picture}(60,120)(-90,-20)
\unitlength 0.8mm
\thicklines


\put(-75,35){\ellipse{4}{8}}
\put(-75,15){\ellipse{4}{8}}
\put(-75,39){\line(1,0){2}}
\put(-75,31){\line(1,0){2}}
\put(-75,19){\line(1,0){2}}
\put(-75,11){\line(1,0){2}}
\put(-73,25){\arc{12}{4.6}{7.9}}
\put(-73,25){\arc{28}{4.6}{5.9}}
\put(-73,25){\arc{28}{6.8}{7.9}}
\put(-55,29){\line(-1,0){2}}
\put(-55,21){\line(-1,0){2}}
\put(-57,33){\arc{8}{1.5}{3}}
\put(-57,17){\arc{8}{3.3}{4.7}}
\put(-55,25){\ellipse{4}{8}}


\put(-55,25){\ellipse{4}{8}}
\put(-55,5){\ellipse{4}{8}}
\put(-55,29){\line(1,0){2}}
\put(-55,21){\line(1,0){2}}
\put(-55,9){\line(1,0){2}}
\put(-55,1){\line(1,0){2}}
\put(-53,15){\arc{12}{4.6}{7.9}}
\put(-53,15){\arc{28}{4.6}{5.9}}
\put(-53,15){\arc{28}{6.8}{7.9}}
\put(-35,19){\line(-1,0){2}}
\put(-35,11){\line(-1,0){2}}
\put(-37,23){\arc{8}{1.5}{3}}
\put(-37,7){\arc{8}{3.3}{4.7}}
\put(-35,15){\ellipse{4}{8}}


\put(-10,25){\ellipse{4}{8}}
\put(-10,5){\ellipse{4}{8}}
\put(-10,29){\line(1,0){2}}
\put(-10,21){\line(1,0){2}}
\put(-10,9){\line(1,0){2}}
\put(-10,1){\line(1,0){2}}
\put(-8,15){\arc{12}{4.6}{7.9}}
\put(-8,15){\arc{28}{4.6}{5.9}}
\put(-8,15){\arc{28}{6.8}{7.9}}
\put(10,19){\line(-1,0){2}}
\put(10,11){\line(-1,0){2}}
\put(8,23){\arc{8}{1.5}{3}}
\put(8,7){\arc{8}{3.3}{4.7}}
\put(10,15){\ellipse{4}{8}}


\put(10,35){\ellipse{4}{8}}
\put(10,15){\ellipse{4}{8}}
\put(10,39){\line(1,0){2}}
\put(10,31){\line(1,0){2}}
\put(10,19){\line(1,0){2}}
\put(10,11){\line(1,0){2}}
\put(12,25){\arc{12}{4.6}{7.9}}
\put(12,25){\arc{28}{4.6}{5.9}}
\put(12,25){\arc{28}{6.8}{7.9}}
\put(30,29){\line(-1,0){2}}
\put(30,21){\line(-1,0){2}}
\put(28,33){\arc{8}{1.5}{3}}
\put(28,17){\arc{8}{3.3}{4.7}}
\put(30,25){\ellipse{4}{8}}

\put(-25,15){$=$}
\put(-32,13){$\sigma$}
\put(33,23){$\sigma$}
\put(-85,34){$\mu$}
\put(2,34){$\mu$}
\put(-20,24){$\nu$}
\put(-85,14){$\nu$}
\put(-20,4){$\rho$}
\put(-65,4){$\rho$}
\put(-43,-10){$\displaystyle{\sum_{\lambda}} C_{\mu \nu}^{\lambda} C_{\lambda \rho}^{\sigma} = \displaystyle{\sum_{\lambda}} C_{\nu \rho}^{\lambda} C_{\mu \lambda}^\sigma$}

\end{picture}
\end{center}
\caption{Associativity of the structure constant.}
\label{fig:associativity}
\end{figure}


\begin{figure}[t]
\begin{center}
\begin{picture}(40,80)(-40,20)
\unitlength 0.7mm
\thicklines

\put(-30,25){\ellipse{5}{10}}
\put(30,25){\ellipse{5}{10}}
\put(-30,20){\line(1,0){6}}
\put(-30,30){\line(1,0){6}}
\put(30,20){\line(-1,0){5}}
\put(30,30){\line(-1,0){5}}

\put(-25,35){\arc{10}{0.7}{1.5}}
\put(-25,15){\arc{10}{4.8}{5.6}}
\put(25,35){\arc{10}{1.5}{2.5}}
\put(25,15){\arc{10}{3.7}{4.8}}
\put(0,18){\arc{50}{3.72}{5.72}}
\put(0,32){\arc{50}{0.59}{2.57}}
\put(0,18){\arc{28}{3.72}{5.72}}
\put(0,34){\arc{30}{0.55}{2.6}}
\put(-40,24){$\nu$}
\put(35,24){$\mu$}
\put(-72,24){$(H \cdot C)_{\nu}^{\, \, \mu}~=$}
\end{picture}
\end{center}
\caption{Handle  creating operator}
\label{fig:handle}
\end{figure}

To write down the genus $g$ partition function in a compact form, we introduce $H^{\lambda}$ and 
the handle operator $(H\cdot C)$;
\begin{align}
&H^{\lambda}:=\sum_{\mu, \nu \in L} \eta^{\mu \nu} C^{\lambda}_{\mu \nu} \,,\\
& (H \cdot C)_{\nu}^{\, \, \mu} = \sum_{\rho \in L} H^{\rho} C^{\mu}_{\nu \rho} \,.
\label{eq:handle}
\end{align}
As shown in  figure \ref{fig:handle}, $H^{\lambda}$ corresponds to a genus one surface 
with a hole with right orientation and $(H \cdot C)_{\nu}^{\, \, \mu}$ corresponds to 
a genus one surface with two holes with left and right orientation.
Then the partition function $Z_g:=\langle 1 \rangle_{g}$ for 
a closed Riemann surface with genus $g$ is expressed   
\begin{align}
Z_g= \sum_{\nu \in L} \{ (H \cdot C)^g \}_{\emptyset}^{\, \, \nu} \, \eta_{\nu \, \emptyset } \,,
\label{eq:gSC}
\end{align}
where we define $(H\cdot C)^g$ as a product of matrices
\begin{align}
\{ (H \cdot C)^g \}_{\mu}^{\, \, \nu}:=\sum_{\mu_1, \cdots \mu_{g-1} \in L} (H \cdot C)_{\mu}^{ \, \, \mu_{1}} (H \cdot C)_{\mu_1}^{\, \, \mu_2} 
\cdots (H \cdot C)_{\mu_{g-2}}^{\, \, \mu_{g-1}} (H \cdot C)_{\mu_{g-1}}^{\, \, \nu}\,.
\end{align}
Since \eqref{eq:gSC} is rewritten as
\begin{align}
Z_g= \mathrm{Tr} (H \cdot C)^{g-1} = \sum_{\mu \in L} \{ (H \cdot C)^{g-1} \}_{\mu}^{\, \, \mu}\label{zgenus}\,,
\end{align}
$Z_g$ is expressed in terms of the roots $y_i$ of the characteristic polynomial of the matrix $H \cdot C$;
\begin{align}
Z_g= \sum_{i} m_i y^{g-1}_i\,,\label{zg-eigenvalue}
\end{align}
with
\begin{align}
\det \left(y I- H \cdot C\right)=\prod_{i} (y-y_i)^{m_i}\,.\label{charact-poly}
\end{align}
Here $I$ stands for the unit matrix of size $|L|$ and 
the integer $m_i$ is the multiplicity of $y_i$.

Finally since all the higher genus correlation functions are obtained by gluing the genus zero two point function
$\eta_{\mu\nu} = C_{\emptyset \mu\nu}$ and structure constants $C^{\lambda}_{\mu\nu}$, 
any correlation function is expressed in terms of  $\eta_{\mu \nu}$  and $S_{\mu\nu}$;
\beq
\langle \mathcal{O}_{\lambda_1} \mathcal{O}_{\lambda_2} \cdots \mathcal{O}_{\lambda_n} \rangle_g
= \sum_{\sigma \in L} \left( \sum_{\mu, \nu \in L} \frac{\eta^{\mu \nu} S_{\mu \sigma} S_{\nu \sigma}   }{S^2_{\emptyset\sigma}} \right)^{g-1} \prod_{i=1}^n  \frac{S_{\lambda_i \sigma}}{S_{\emptyset\sigma}}\,.
\eeq

%
\section{Hall-Littlewood polynomial}

The Hall-Littlewood polynomial $P_{\lambda}(x,t)$ is an important family of symmetric polynomials,
which is regarded as a deformation of the Schur polynomial $s_{\lambda}(x)$. 
Let $\lambda$ be a partition $\lambda = (\lambda_1 \geq \lambda_2 \geq \cdots \geq \lambda_N)$ of length (at most) $N$.
We introduce the following polynomial with $N$ variables $x = (x_1, x_2, \cdots, x_N)$;
\begin{align}
R_{\lambda}(x,t)= \sum_{\omega \in \mathfrak{S}_N}   \left(x^{\lambda_1}_{\omega(1)} \cdots x^{\lambda_N}_{\omega(N)} \prod_{a < b}^{N}
\frac{x_{\omega(a)}-t x_{\omega(b)} }{x_{\omega(a)} -x_{\omega(b)}}\right),
\end{align}
where $\mathfrak{S}_N$ is the symmetric group of $N$ objects and $t$ is an indeterminate (parameter). 
Then we can define the Hall-Littlewood polynomial by
\beq
P_{\lambda}(x,t)=\frac{1}{v_{\lambda}(t)} R_{\lambda}(x,t), \qquad
v_{\lambda}(t):=\prod_{i=0}^{\infty} \prod_{j=1}^{m_i(\lambda)} \frac{1-t}{1-t^j}.
\eeq
Then $P_{\lambda}(x,t)$ gives a $\Z[t] $-basis of the ring of the symmetric polynomials 
$\Z[t] [x_1, x_2, \cdots x_N]^{\mathfrak{S}_N}$.
Note that $P_{\lambda}(x,t)$ provides  interpolation between the Schur polynomial $s_\lambda(x)$ and  
the symmetric monomial  $m_\lambda(x)$ 
\begin{align}
P_{\lambda}(x,0) = s_\lambda(x), \qquad P_{\lambda}(x,1) = m_\lambda(x).
\end{align}
When one changes bases from the  Hall-Littlewood polynomials $P_{\lambda}(x,t)$ to 
the Schur polynomials $s_\lambda (x)$, 
its efffect is realised as 
a matrix $K_{\lambda \mu} $;
\begin{align}
s_\lambda (x) = \sum_{ |\mu|=|\lambda|} K_{\lambda \mu} (t) P_{\mu}(x,t)\,.   
\end{align}
This matrix has triangular form with respect to the dominance semi-ordering of partitions. 
$K_{\lambda \mu} (t)$ is called the 
Kostka polynomial ($K_{\lambda \mu} (1) = K_{\lambda \mu}$ are the Kostka numbers)
and is ubiquitous in representation theories 
and combinatorics. One of the most important properties of $K_{\lambda \mu} (t)$
is that all the coefficients are non-negative integer, which gives us  an interpretation 
of  dimensions of appropriate modules.

In our method of computing the structure constants of $U(N)$ equivariant Verlinde algebra with level $\kappa$,
after substituting a root of the Bethe ansatz equation to $x$,
we have to reduce the Hall-Littlewood polynomials $P_{\lambda}(x,t) $ for any partition $\lambda$ of length $N$
to a linear combination of $P_{\mu}(x,t)$, where $\mu$ runs only in $\mathcal{P}_{N,\kappa}$.
We make use of the relations derived in \cite{Korff:2013rsa} for this purpose.
Mathematically these relations generate an ideal $\mathcal{I}_{N,\kappa}$ in the ring of Hall-Littlewood polynomials.
This means that we identify the equivariant Verlinde algebra with a quotient of the ring of Hall-Littlewood polynomials by $\mathcal{I}_{N,\kappa}$.
This algorithm does work, since any $\lambda$ regarded as a weight vector of $\mathfrak{gl}(N)$,
can be transformed into $\mathcal{P}_{N,\kappa}$ by the affine Weyl group $\widetilde{\mathfrak{S}}_{N,\kappa}$
with level $\kappa$. 
In this sense $\mathcal{P}_{N,\kappa}$ is a fundamental domain for $\widetilde{\mathfrak{S}}_{N,\kappa}$.
The group $\widetilde{\mathfrak{S}}_{N,\kappa}$ is generated by $\sigma_i~(1 \leq i \leq N-1), \sigma_0$ and $\tau$.
The (right) action on a weight $\lambda$ is defined by
\begin{align}
&\lambda \cdot \sigma_i:=(\lambda_{1}, \lambda_2, \cdots, \lambda_{i+1},\lambda_{i}, \cdots, \lambda_{N})\,, \\
&\lambda \cdot \sigma_0:=(\lambda_{N}+\kappa, \lambda_2, \cdots, \lambda_{1}-\kappa)\,, \\
&\lambda \cdot \tau:=(\lambda_{N}+\kappa, \lambda_1, \lambda_2, \cdots, \lambda_{N-1})\,.
\end{align}
If we substitute a Bethe root $x =(x_1,\cdots, x_N)$, we have the following identities\footnote{We use $\equiv$ to emphasize 
the equality on the space of Bethe roots.};
\begin{align}
&R_{\lambda }(x,t) \equiv R_{\lambda \cdot \tau }(x,t), 
\label{eq:AppB2} \\
&R_{\lambda \cdot \sigma_i }(x,t) \equiv t R_{\lambda}(x,t), \quad \lambda_i -\lambda_{i+1}=1,  
\label{eq:AppB3} \\
&R_{\lambda \cdot \sigma_i }(x,t) \equiv t R_{\lambda}(x,t)+(t-1) 
R_{(\lambda_1, \cdots, \lambda_{i-1}, \lambda_{i}+1,\lambda_{i}+1, \lambda_{i+2}, \cdots,\lambda_N)}(x,t), \quad \lambda_i -\lambda_{i+1}=2,  
\label{eq:AppB4} \\
&R_{\lambda \cdot \sigma_0 }(x,t) \equiv t R_{\lambda}(x,t)-R_{(\lambda_1+1, \lambda_2, \cdots, \lambda_{N-1}, \lambda_{N}-1)}(x,t)
+tR_{(\lambda_{N}-1+\kappa, \lambda_2, \cdots, \lambda_{N-1}, \lambda_{1}+1-\kappa)}(x,t),
\label{eq:AppB5} 
\end{align}
and
\beq
P_{\lambda}(x,t) \equiv P_{\tilde{\lambda}}(x,t), \quad \lambda \in \mathcal{P}_{N, \kappa},
\label{eq:AppB1} 
\eeq
where $\tilde{\lambda}$ is obtained by deleting all the rows of size $\kappa$.


\section{Couplings $C^{(r)}_{\mu\nu\rho}$ for generic R-charge $r$}

In this appendix, we summarize properties of 
the three point function $C^{(r)}_{\mu\nu\rho}$ defined by Hall-Littlewood polynomials 
in the case of generic R-charge $r$.

First we put $r=0$ for simplicity and study 
metrics $\eta^{(r=0)}_{\mu\nu}$. 
The structure constants $C_{\mu\nu}^{\lambda}$ are defined from the product of Hall-Littlewood polynomials in \eqref{eq:deffusion},
which are independent of $R$-charge $r$.
The metric in this case $r=0$ is obtained in the paper \cite{Korff:2013rsa}
\begin{eqnarray}
&&\eta^{(r=0)}_{\mu\nu}:=\frac{\delta_{\mu\nu^{\ast}}}{b_{\mu}(t)}\,,\label{eta0}
\end{eqnarray}
where $b_{\mu}(t)$ is defined by \eqref{bmu}.
The structure constants $C_{\mu\nu}^{\lambda}$ satisfy the following basic properties:

\begin{enumerate}
\item Symmetric property
\beq
C_{\mu\nu}^{\lambda}=
C_{\nu\mu}^{\lambda}\,,\qquad
\eta^{(r=0)}_{\mu\nu}=
\eta^{(r=0)}_{\nu\mu}\,.
\eeq
\item Existence of the unit operator "$1$" corresponding to $\emptyset = (\kappa, \cdots, \kappa)$
\beq
C_{\emptyset \mu}^{\nu}=\delta_{\mu}{}^{\nu}\,.
\eeq
\item Associativity relation
\beq
\sum_{\alpha ,\beta}C_{\mu_1\mu_2}^{\alpha}
\eta^{(r=0)}_{\alpha\beta}C_{\mu_3\lambda}^{\beta}
=\sum_{\alpha ,\beta}C_{\mu_1\mu_3}^{\alpha}
\eta^{(r=0)}_{\alpha\beta}C_{\mu_2\lambda}^{\beta}\,.
\eeq
\end{enumerate}

We can also define couplings $C^{(r=0)}_{\mu\nu\rho}$ with three subscripts by
\begin{eqnarray}
&&C^{(r=0)}_{\mu\nu\rho}:=\sum_{\lambda}C_{\mu\nu}^{\lambda}\eta^{(r=0)}_{\lambda\rho}\,.\label{sym3}
\end{eqnarray}
Then they are totally symmetric under the exchange of indices;
\begin{eqnarray}
&&C^{(r=0)}_{\mu_1\mu_2\mu_3}=
C^{(r=0)}_{\mu_2\mu_1\mu_3}\,,\qquad
C^{(r=0)}_{\mu_1\mu_2\mu_3}=
C^{(r=0)}_{\mu_1\mu_3\mu_2}\,,
\end{eqnarray}
which can be derived from the existence of the unit operator and  the associativity relation above.

Next we shall consider couplings $C^{(r)}_{\mu\nu\rho}$ for integral $R$-charge $r$.
In order to define them (see the formula \eqref{eq:corCSadj}), 
we need the expansion of the product $\prod_{a,b} (1-tx_ax_b^{-1})^r$ by the Hall-Littlewood polynomials;
\begin{eqnarray}
&&\prod_{a,b=1}^N
(1-tx_ax_b^{-1})^r\equiv
(1-t)^{rN} \sum_{\lambda \in \mathcal{P}_{N, \kappa} }g^{(r)}_{\lambda}(t)P_{\lambda}(x,t)\,.\label{expansion}
\end{eqnarray}
Then the couplings 
$C^{(r)}_{\mu\nu\rho}$ and metrics $\eta^{(r)}_{\mu\nu}$ are 
related to the structure constants of the Hall-Littlewood polynomials
as follows;
\begin{eqnarray}
&&C^{(r)}_{\mu\nu\rho}(t):=(1-t)^{rN} 
\sum_{\lambda ,\alpha ,\beta} g^{(r)}_{\lambda}(t)
C_{\mu\nu}^{\alpha}\eta^{(r=0)}_{\alpha\beta} C_{\rho\lambda}^{\beta}
=\sum_{\lambda}C_{\mu\nu}^{\lambda} \eta^{(r)}_{\lambda\rho}
\,,\label{r-sym3}\\
&&\eta^{(r)}_{\mu\nu}:=C^{(r)}_{ \emptyset \mu\nu}\,,\qquad
C^{(r=0)}_{\lambda\mu\nu}:=C_{\lambda\mu\nu} \label{eta-r}\,.
\end{eqnarray}
Note that the $R$-charge dependence appears only through $(1-t)^{rN} g^{(r)}_{\lambda}(t)$
which determines the metric $\eta^{(r)}_{\lambda\rho}$.
We can prove the fusion couplings $C^{(r)}_{\mu\nu\rho}$ are invariant under the exchange of the subscripts
\begin{eqnarray}
&&C^{(r)}_{\mu_1\mu_2\mu_3}=C^{(r)}_{\mu_2\mu_1\mu_3}\,,\qquad
C^{(r)}_{\mu_1\mu_2\mu_3}=C^{(r)}_{\mu_1\mu_3\mu_2}\,,
\end{eqnarray}
where the first relation is proved by using the symmetry of 
$C_{\mu_1\mu_2}^{\alpha}$ and the definition of 
$C^{(r)}_{\mu_1\mu_2\mu_3}$.
The second relation can be shown by using the associativity relation.
As a result of this symmetry, the metric 
$\eta^{(r)}_{\mu\nu}=C^{(r)}_{\emptyset\mu\nu}$ is symmetric 
$\eta^{(r)}_{\mu\nu}=\eta^{(r)}_{\nu\mu}$ as it should be.
We can also show the associativity of the couplings $C^{(r)}_{\mu\nu\rho}$
\beq
\sum_{\alpha ,\beta}C_{\mu_1\mu_2}^{\alpha}\eta^{(r)}_{\alpha\beta}C_{\mu_3\lambda}^{\beta}
=
\sum_{\alpha ,\beta}C_{\mu_1\mu_3}^{\alpha}\eta^{(r)}_{\alpha\beta}C_{\mu_2\lambda}^{\beta} \,.
\eeq




\begin{thebibliography}{99}

\bibitem{Pestun:2016zxk} 
  V.~Pestun {\it et al.},
  ``Localization techniques in quantum field theories,''
  J.\ Phys.\ A {\bf 50}, no. 44, 440301 (2017)
  [arXiv:1608.02952 [hep-th]].

\bibitem{Moore:1997dj} 
  G.~W.~Moore, N.~Nekrasov and S.~Shatashvili,
  ``Integrating over Higgs branches,''
  Commun.\ Math.\ Phys.\  {\bf 209}, 97 (2000)
  [hep-th/9712241].

\bibitem{Nekrasov:2009uh} 
  N.~A.~Nekrasov and S.~L.~Shatashvili,
  ``Supersymmetric vacua and Bethe ansatz,''
  Nucl.\ Phys.\ Proc.\ Suppl.\  {\bf 192-193}, 91 (2009)
  [arXiv:0901.4744 [hep-th]].

\bibitem{Nekrasov:2009ui} 
  N.~A.~Nekrasov and S.~L.~Shatashvili,
  ``Quantum integrability and supersymmetric vacua,''
  Prog.\ Theor.\ Phys.\ Suppl.\  {\bf 177}, 105 (2009)
  [arXiv:0901.4748 [hep-th]].

\bibitem{Nekrasov:2009rc} 
  N.~A.~Nekrasov and S.~L.~Shatashvili,
  ``Quantization of Integrable Systems and Four Dimensional Gauge Theories,''
  arXiv:0908.4052 [hep-th].

\bibitem{Nekrasov:2014xaa} 
  N.~A.~Nekrasov and S.~L.~Shatashvili,
  ``Bethe/Gauge correspondence on curved spaces,''
  JHEP {\bf 1501}, 100 (2015)
  [arXiv:1405.6046 [hep-th]].

\bibitem{Alday:2009aq} 
  L.~F.~Alday, D.~Gaiotto and Y.~Tachikawa,
  ``Liouville Correlation Functions from Four-dimensional Gauge Theories,''
  Lett.\ Math.\ Phys.\  {\bf 91}, 167 (2010)
  [arXiv:0906.3219 [hep-th]].

\bibitem{Wyllard:2009hg} 
  N.~Wyllard,
  ``A(N-1) conformal Toda field theory correlation functions from conformal N = 2 SU(N) quiver gauge theories,''
  JHEP {\bf 0911}, 002 (2009)
  [arXiv:0907.2189 [hep-th]].
  
\bibitem{Terashima:2011qi} 
  Y.~Terashima and M.~Yamazaki,
  ``SL(2,R) Chern-Simons, Liouville, and Gauge Theory on Duality Walls,''
  JHEP {\bf 1108}, 135 (2011)
  [arXiv:1103.5748 [hep-th]].
  
\bibitem{Terashima:2011xe} 
  Y.~Terashima and M.~Yamazaki,
  ``Semiclassical Analysis of the 3d/3d Relation,''
  Phys.\ Rev.\ D {\bf 88}, no. 2, 026011 (2013)
  [arXiv:1106.3066 [hep-th]].
  
\bibitem{Dimofte:2011ju} 
  T.~Dimofte, D.~Gaiotto and S.~Gukov,
  ``Gauge Theories Labelled by Three-Manifolds,''
  Commun.\ Math.\ Phys.\  {\bf 325}, 367 (2014)
  [arXiv:1108.4389 [hep-th]].

\bibitem{Dimofte:2011py} 
  T.~Dimofte, D.~Gaiotto and S.~Gukov,
  ``3-Manifolds and 3d Indices,''
  Adv.\ Theor.\ Math.\ Phys.\  {\bf 17}, no. 5, 975 (2013)
  [arXiv:1112.5179 [hep-th]].
   
\bibitem{Cordova:2013cea} 
  C.~Cordova and D.~L.~Jafferis,
  ``Complex Chern-Simons from M5-branes on the Squashed Three-Sphere,''
  JHEP {\bf 1711}, 119 (2017)
  [arXiv:1305.2891 [hep-th]].

\bibitem{Lee:2013ida} 
  S.~Lee and M.~Yamazaki,
  ``3d Chern-Simons Theory from M5-branes,''
  JHEP {\bf 1312}, 035 (2013)
  [arXiv:1305.2429 [hep-th]].

\bibitem{Dimofte:2014zga} 
  T.~Dimofte,
  ``Complex Chern-Simons Theory at Level k via the 3d-3d Correspondence,''
  Commun.\ Math.\ Phys.\  {\bf 339}, no. 2, 619 (2015)
  [arXiv:1409.0857 [hep-th]].

\bibitem{Dimofte:2016pua} 
  T.~Dimofte,
  ``Perturbative and nonperturbative aspects of complex Chern-Simons theory,''
  J.\ Phys.\ A {\bf 50}, no. 44, 443009 (2017)
  [arXiv:1608.02961 [hep-th]].
  
\bibitem{Witten:1988hf} 
  E.~Witten,
  ``Quantum Field Theory and the Jones Polynomial,''
  Commun.\ Math.\ Phys.\  {\bf 121}, 351 (1989).


\bibitem{Verlinde:1988sn} 
  E.~P.~Verlinde,
  ``Fusion Rules and Modular Transformations in 2D Conformal Field Theory,''
  Nucl.\ Phys.\ B {\bf 300}, 360 (1988).
  
\bibitem{Gepner:1990gr} 
  D.~Gepner,
  ``Fusion rings and geometry,''
  Commun.\ Math.\ Phys.\  {\bf 141}, 381 (1991).

\bibitem{Intriligator:1991an} 
  K.~A.~Intriligator,
  ``Fusion residues,''
  Mod.\ Phys.\ Lett.\ A {\bf 6}, 3543 (1991)
  [hep-th/9108005].

\bibitem{Witten:1993xi} 
  E.~Witten,
  ``The Verlinde algebra and the cohomology of the Grassmannian,''
  In *Cambridge 1993, Geometry, topology, and physics* 357-422
  [hep-th/9312104].
  
\bibitem{Gukov:2015sna} 
  S.~Gukov and D.~Pei,
  ``Equivariant Verlinde formula from fivebranes and vortices,''
  Commun.\ Math.\ Phys.\  {\bf 355}, no. 1, 1 (2017)
  [arXiv:1501.01310 [hep-th]].
  
    
\bibitem{Andersen:2016hoj} 
  J.~E.~Andersen, S.~Gukov and D.~Pei,
  ``The Verlinde formula for Higgs bundles,''
  arXiv:1608.01761 [math.AG].
  
\bibitem{Gukov:2016lki} 
  S.~Gukov, D.~Pei, W.~Yan and K.~Ye,
  ``Equivariant Verlinde Algebra from Superconformal Index and Argyres-Seiberg Duality,''
  Commun.\ Math.\ Phys.\  {\bf 357}, no. 3, 1215 (2018)
  [arXiv:1605.06528 [hep-th]].
  
\bibitem{Gadde:2009kb} 
  A.~Gadde, E.~Pomoni, L.~Rastelli and S.~S.~Razamat,
  ``S-duality and 2d Topological QFT,''
  JHEP {\bf 1003}, 032 (2010)
  [arXiv:0910.2225 [hep-th]].
  
\bibitem{Benini:2011nc} 
  F.~Benini, T.~Nishioka and M.~Yamazaki,
  ``4d Index to 3d Index and 2d TQFT,''
  Phys.\ Rev.\ D {\bf 86}, 065015 (2012)
  [arXiv:1109.0283 [hep-th]].  

\bibitem{Ohta:2012ev} 
  K.~Ohta and Y.~Yoshida,
  ``Non-Abelian Localization for Supersymmetric Yang-Mills-Chern-Simons Theories on Seifert Manifold,''
  Phys.\ Rev.\ D {\bf 86}, 105018 (2012)
  [arXiv:1205.0046 [hep-th]].

\bibitem{Benini:2015noa} 
  F.~Benini and A.~Zaffaroni,
  ``A topologically twisted index for three-dimensional supersymmetric theories,''
  JHEP {\bf 1507}, 127 (2015)
  [arXiv:1504.03698 [hep-th]].

\bibitem{Benini:2016hjo} 
  F.~Benini and A.~Zaffaroni,
  ``Supersymmetric partition functions on Riemann surfaces,''
  Proc.\ Symp.\ Pure Math.\  {\bf 96}, 13 (2017)
  [arXiv:1605.06120 [hep-th]].

\bibitem{Closset:2016arn} 
  C.~Closset and H.~Kim,
  ``Comments on twisted indices in 3d supersymmetric gauge theories,''
  JHEP {\bf 1608}, 059 (2016)
  [arXiv:1605.06531 [hep-th]].
   
  
    
\bibitem{Okuda:2012nx} 
  S.~Okuda and Y.~Yoshida,
  ``G/G gauged WZW model and Bethe Ansatz for the phase model,''
  JHEP {\bf 1211}, 146 (2012)
  [arXiv:1209.3800 [hep-th]].

\bibitem{Okuda:2013fea} 
  S.~Okuda and Y.~Yoshida,
  ``G/G gauged WZW-matter model, Bethe Ansatz for q-boson model and Commutative Frobenius algebra,''
  JHEP {\bf 1403}, 003 (2014)
  [arXiv:1308.4608 [hep-th]].

\bibitem{Okuda:2015yea} 
  S.~Okuda and Y.~Yoshida,
  ``Gauge/Bethe correspondence on $S^1 \times \Sigma_h$ and index over moduli space,''
 [arXiv:1501.03469 [hep-th].


\bibitem{Benini:2013xpa} 
  F.~Benini, R.~Eager, K.~Hori and Y.~Tachikawa,
  ``Elliptic Genera of 2d ${\mathcal{N}}$ = 2 Gauge Theories,''
  Commun.\ Math.\ Phys.\  {\bf 333}, no. 3, 1241 (2015)
  [arXiv:1308.4896 [hep-th]].

 
\bibitem{Korff:2013rsa} 
  C.~Korff,
  ``Cylindric versions of specialised macdonald functions and a deformed Verlinde algebra,''
  Commun.\ Math.\ Phys.\  {\bf 318}, 173 (2013)
  [arXiv:1110.6356 [math-ph]].



\bibitem{Alday:2013rs} 
  L.~F.~Alday, M.~Bullimore and M.~Fluder,
  ``On S-duality of the Superconformal Index on Lens Spaces and 2d TQFT,''
  JHEP {\bf 1305}, 122 (2013)
  [arXiv:1301.7486 [hep-th]].
 
\bibitem{Razamat:2013jxa} 
  S.~S.~Razamat and M.~Yamazaki,
  ``S-duality and the N=2 Lens Space Index,''
  JHEP {\bf 1310}, 048 (2013)
  [arXiv:1306.1543 [hep-th]].
 


\bibitem{Romelsberger:2005eg} 
  C.~Romelsberger,
  ``Counting chiral primaries in N = 1, d=4 superconformal field theories,''
  Nucl.\ Phys.\ B {\bf 747}, 329 (2006)
  [hep-th/0510060].

\bibitem{Kinney:2005ej} 
  J.~Kinney, J.~M.~Maldacena, S.~Minwalla and S.~Raju,
  ``An Index for 4 dimensional super conformal theories,''
  Commun.\ Math.\ Phys.\  {\bf 275}, 209 (2007)
  [hep-th/0510251].
  
\bibitem{Gadde:2011uv} 
  A.~Gadde, L.~Rastelli, S.~S.~Razamat and W.~Yan,
  ``Gauge Theories and Macdonald Polynomials,''
  Commun.\ Math.\ Phys.\  {\bf 319}, 147 (2013)
  [arXiv:1110.3740 [hep-th]].

\bibitem{Macdonald:book} 
I.~G.~Macdonald, 
``Symmetric functions and Hall Polynomials, Second Edition (1995),'' 
OXFORD UNIVERSTY PRESS. 
 
\bibitem{Nakanishi:1990hj} 
  T.~Nakanishi and A.~Tsuchiya,
  ``Level rank duality of WZW models in conformal field theory,''
  Commun.\ Math.\ Phys.\  {\bf 144}, 351 (1992).

\bibitem{Naculich:1990pa} 
  S.~G.~Naculich, H.~A.~Riggs and H.~J.~Schnitzer,
  ``Group Level Duality in {WZW} Models and {Chern-Simons} Theory,''
  Phys.\ Lett.\ B {\bf 246}, 417 (1990).

\bibitem{Mlawer:1990uv} 
  E.~J.~Mlawer, S.~G.~Naculich, H.~A.~Riggs and H.~J.~Schnitzer,
  ``Group level duality of WZW fusion coefficients and Chern-Simons link observables,''
  Nucl.\ Phys.\ B {\bf 352}, 863 (1991).

\bibitem{Naculich:2007nc} 
  S.~G.~Naculich and H.~J.~Schnitzer,
  ``Level-rank duality of the U(N) WZW model, Chern-Simons theory, and 2-D qYM theory,''
  JHEP {\bf 0706}, 023 (2007)
  [hep-th/0703089 [HEP-TH]].

\bibitem{Hsin:2016blu} 
  P.~S.~Hsin and N.~Seiberg,
  ``Level/rank Duality and Chern-Simons-Matter Theories,''
  JHEP {\bf 1609}, 095 (2016)
  [arXiv:1607.07457 [hep-th]



\end{thebibliography}
\end{document}